\documentclass[aps,graphicx,twocolumn,epsfig,showpacs]{revtex4}

\usepackage{epsfig}
\usepackage{amsmath}
\usepackage{amsfonts}
\usepackage{graphicx}

\newcommand{\ket}[1]{|{#1}\rangle}                     % | >
\newcommand{\bra}[1]{\langle{#1}|} 
\newcommand{\braket}[2]{\langle{#1}|{#2}\rangle}

\newcommand{\kk}{\mathbf{k}}
\newcommand{\KK}{\mathbf{K}}
\newcommand{\PP}{\mathbf{P}}
\newcommand{\qq}{\mathbf{q}}
\newcommand{\OO}{\mathbf{0}}
\newcommand{\ve}{\mathbf{e}}

\newcommand{\be}{\begin{equation}}
\newcommand{\ee}{\end{equation}}
\newcommand{\bea}{\begin{eqnarray}}
\newcommand{\eea}{\end{eqnarray}}

\begin{document}

\author{Christian Trefzger}
\author{Yvan Castin} 
\affiliation{Laboratoire Kastler Brossel, \'Ecole Normale Sup\'erieure and CNRS, UPMC, 24 rue Lhomond, 75231 Paris, France}

\title{An impurity in a Fermi sea on a narrow Feshbach resonance:
A variational study of the polaronic and dimeronic branches}

\begin{abstract}
We study the problem of a single impurity of mass $M$ immersed in a Fermi sea of particles of mass $m$. The impurity and the fermions interact through a $s$-wave narrow Feshbach resonance, 
so that the Feshbach length $R_*$ naturally appears in the system. We use simple variational ansatz, limited to at most one pair of particle-hole excitations of the Fermi sea and we determine
for the polaronic and dimeronic branches the phase diagram between absolute ground state, local minimum, thermodynamically unstable regions (with negative effective mass), and regions of complex energies (with negative imaginary part).
We also determine the closed channel population which is experimentally accessible. 
Finally we identify a non-trivial weakly attractive limit where analytical results can be obtained, in particular for the crossing point between the polaronic and dimeronic 
energy branches.  
\end{abstract}


\maketitle

\section{Introduction}

The physics of strongly interacting Fermi gases is now experiencing a rapid development. Experimentalists have succeeded in obtaining highly degenerate 
and strongly interacting 
samples, providing new perspectives in the study of Fermi superfluidity, including BCS-BEC crossover: 
Thanks to magnetic Feshbach resonance techniques one can indeed control at will the interaction strength and even 
reach the unitary limit, where interaction effects are maximal \cite{ketterle,Giorgini}. 
Furthermore very precise 
measurements of the many body properties of these systems are at reach: Recently the equation of state of 
the gas at the unitary limit or within the whole BEC-BCS crossover was measured with an uncertainty 
at a few percent level \cite{Nascimbene,Zwierlein,VanHoucke}.

Of particular interest is the case of spin imbalanced Fermi gases, that is with a larger number of spin up 
particles ($N_\uparrow\gg N_\downarrow$). In this spin polarized regime, the usual Cooper pairing scenario no longer applies, and 
this raises the question of the possible phases of the system. Apart from the FFLO phase \cite{Combescot_Mora}, and more 
exotic phases \cite{GSarma}, it was proposed recently that, 
the minority atoms dressed by the Fermi sea of the majority atoms form a weakly interacting gas of quasi-particles called 
polarons \cite{Chevy,Lobo}. 

Whereas the basic properties of these polarons, such as their binding energy with the Fermi sea, and their effective mass,
are  well understood in the limit where the majority atoms interact with the minority atoms resonantly on a $s$-wave wide Feshbach resonance \cite{Lobo,Svistunov,Combescot,Combescot_bs},
their properties have not been studied yet when the Feshbach resonance becomes narrow.
A narrow Feshbach resonance is characterized by a length $R_*>0$, 
called the Feshbach length, which vanishes in the broad resonance
limit $R_*=0$ \cite{Petrov}.
Therefore the presence of a positive $R_*$ introduces a new length in the system, in addition to the $s$-wave scattering length $a$, which may be of physical importance depending on how it compares
with the other physical parameters. It is then natural to ask  whether the impurity problem on a narrow Feshbach resonance 
presents new physics. Of particular interest is the case of fermionic mixtures of $^6$Li and $^{40}$K, which in the last decades of experiments have been extensively realized.
It turns out that all the Feshbach resonances in such mixtures are narrow~\cite{Narrow}, characterized by a value of the Feshbach length $R_* \gtrsim 100{\rm nm}$,
which is much larger than the typical van der Waals length $b\simeq2{\rm nm}$. It is therefore expected that a fermionic mixture of $^6$Li and $^{40}$K
will present some physical effects due to the narrowness of the Feshbach resonances, as recently demonstrated in the experiment of \cite{RGrimm}.

%In this paper we study the single polaron problem, that is a single impurity in a Fermi sea, on a narrow Feshbach resonance. 
In this paper we study the single impurity problem on a narrow Feshbach resonance, that is a single impurity interacting through a $s$-wave narrow Feshbach resonance 
with a Fermi sea. As shown in \cite{Svistunov}  on a broad Feshbach resonance this problem has two energy branches, one that continuously connects to the free impurity
in the weakly attractive limit ($a\to 0^-$), that we call the {\sl polaronic} branch, and the other one that continuously connects to the free impurity-atom dimer for $a\to 0^+$,
which we call the {\sl dimeronic} branch.
We find that the scattering length value at the crossing point between the two energy branches strongly depends on
the Feshbach length $R_*$, as well as on the mass ratio between the impurity and one of the fermions.  In particular, for all the mass ratios considered in this work we find that as $R_*$
increases, the domain of the polaronic ground state shrinks and shifts progressively towards the weakly attractive region. For any negative scattering length, we find that for a
sufficiently large $R_*$ the ground state is dimeronic.
Furthermore, we identify a non-trivial weakly attractive limit, $a\to 0^-$ with $R_* a$ fixed,  which allows us to solve in an asymptotic expansion the single impurity problem and to obtain the crossing point
between the polaronic and dimeronic branches.

The paper is organized as follows. After introducing the narrow Feshbach resonance on a two-channel model in section~\ref{sec:themodel}, we discuss the variational
ansatz in section~\ref{sec:theansatz}: We present the physical motivation behind the ansatz, we show that the problem falls naturally into the
well studied class of a discrete state coupled to a continuum, and we derive the variational eigenvalue equations. In section~\ref{sec:properties} we numerically
solve the eigenvalue equations and we calculate the properties of the system, including the crossing point, the effective mass and the closed-channel molecule population.
In section~\ref{sec:ariantwalapz} we develop a non-trivial weakly attractive limit that we can treat analytically in a perturbative way. We then show that the analytical solution compare remarkably well with the 
numerical results of section~\ref{sec:properties}, even for parameter values that are not in the asymptotic region. We conclude in section~\ref{sec:conclusion}.

\section{Two-channel model Hamiltonian}
\label{sec:themodel}

We consider a Fermi gas of same-spin-state particles of mass $m$ perturbed by the presence of an impurity, 
that is an extra, distinguishable particle of mass $M$.
We assume that there is no interaction among the fermions, and that the impurity interacts with each fermion
resonantly on a $s$-wave narrow Feshbach resonance.
We use an usual two-channel model to describe this resonant interaction 
\cite{Hussein,Holland,KoehlerBurnett,Koehler,Lee,Gurarie,ondep,Mora_3corps,Kokkelmans,Castin01}: 
The particles exist either in
the form of atoms in the so-called open channel  (annihilation operators $\hat{u}$ and $\hat{d}$ 
for the fermions and the impurity, respectively),
or in the form of a tightly bound molecule of a fermion with the impurity (annihilation operator $\hat{b}$), 
in the closed channel.
We shall take the limit of an infinitely narrow Feshbach resonance \cite{contact_conditions}, neglecting the direct
interaction between the atoms in the open channel,
and including only the coherent interconversion
of the closed-channel molecule into a pair of atoms. The resulting second-quantized Hamiltonian for the system
enclosed in a cubic quantization volume $V$ with periodic boundary conditions is then
\begin{multline}
\hat{H} = \sum_{\kk} \left[\varepsilon_\kk \hat{u}_\kk^\dag \hat{u}_\kk + E_\kk \hat{d}_\kk^\dag \hat{d}_\kk 
+\left(\frac{\varepsilon_\kk}{1+r} + E_\mathrm{mol}\right) \hat{b}_\kk^\dag \hat{b}_\kk \right]\\
+ \frac{\Lambda}{\sqrt{V}} \sum_{\kk,\kk^\prime} \chi(\kk_{\rm rel})
(\hat{b}_{\kk+\kk'}^\dag \hat{u}_{\kk} \hat{d}_{\kk^\prime} + \mathrm{h.c.})\,,
\label{eq:hamil}
\end{multline}
where $\hat{u}_\kk^\dag$ creates a fermion with wave vector $\kk$, $\hat{d}_\kk^\dag$ creates an impurity atom
with wave vector $\kk$, and $b_\kk^\dag$ creates a closed-channel molecule of center-of-mass wave vector $\kk$.
The $\hat{u}_\kk$'s obey the usual fermionic anticommutation relations,
$\{\hat{u}_\kk,\hat{u}_{\kk^\prime}^\dag\}=\delta_{\kk\kk^\prime}$. 
Since there is at most one impurity atom and one closed-channel molecule in our system, their exchange symmetry is irrelevant.
We can imagine for example that the impurity is a fermion, and that the closed-channel molecule is a boson,
so that $\{\hat{d}_\kk,\hat{d}_{\kk^\prime}^\dag\}=\delta_{\kk\kk^\prime}$ and $[\hat{b}_\kk,\hat{b}_{\kk^\prime}^\dag]=\delta_{\kk\kk^\prime}$.
The single-particle kinetic energies are $\varepsilon_\kk = \frac{\hbar^2\kk^2}{2m}$ for the fermions, 
$E_\kk = \frac{\hbar^2\kk^2}{2M}$ for the impurity, and 
\be
r=\frac{M}{m},
\ee
is the impurity-to-fermion mass ratio. 
$E_\mathrm{mol}$ is the internal energy of the closed-channel molecule.
The last contribution in Eq.~(\ref{eq:hamil}) corresponds to 
the coupling between open and closed channels with a coupling constant $\Lambda$.
It contains the cut-off function $\chi$ depending on the relative wave vector
for a fermion of wave vector $\kk$ and the impurity of wave vector $\kk'$:
\be
\kk_{\rm rel} = \mu \left(\frac{\kk}{m}-\frac{\kk'}{M}\right),
\ee
where $\mu=mM/(m+M)$ is the reduced mass.
$\chi$ is a real and isotropic function of $\kk_{\rm rel}$. Its specific momentum dependence is
here irrelevant, provided that $\chi(\kk_{\rm rel})$ tends to unity for $k_{\rm rel}\to 0$,
and tends rapidly to zero  for $k_{\rm rel}\to +\infty$,
so as to make the Hamiltonian problem well defined.

In practice, in all the explicit calculations to come, we shall take the infinite cut-off limit, that is the limit
where the momentum width of the function $\chi$ tends to infinity (so that $\chi(\kk_{\rm rel})$ tends to unity at all 
fixed $\kk_{\rm rel}$).  In this limit, $\Lambda$ is fixed but $E_{\rm mol}$ is continuously
adjusted so that the interaction is finally characterized
by the usual two physical parameters, the $s$-wave scattering length
$a$ (of arbitrary sign) and the always positive Feshbach length $R_*$ \cite{Petrov}.
In terms of the effective coupling constant,
\be
\label{eq:g}
g = \frac{2\pi\hbar^2 a}{\mu},
\ee
the internal energy of the molecule is adjusted as (see e.g.\ \cite{Castin01}):
\be
\label{eq:Emol}
\frac{E_\mathrm{mol}}{\Lambda^2} = -\frac{1}{g} + \int\frac{d^3k}{(2\pi)^3} \chi^2(\kk) \frac{2\mu}{\hbar^2k^2},
\ee
and the interchannel coupling amplitude is related to the Feshbach length by 
\be
\label{eq:Rstar}
R_*=\frac{\pi\hbar^4}{\Lambda^2\mu^2}.
\ee
Interestingly, $R_*$ can be related to experimental parameters characterizing the Feshbach resonance,
in particular its magnetic width $\Delta B$ \cite{link}.
In the present infinite cut-off limit, note that the $s$-wave scattering amplitude
between a fermion and the impurity in the center-of-mass frame is simply
\be
\label{eq:amplitude}
f_{k_{\rm rel}}=-\frac{1}{\frac{1}{a}+i k_{\rm rel}+k_{\rm rel}^2 R_*},
\ee
so that the effective range is $r_e=-2 R_*$. By inspection of the poles of the scattering amplitude
one readily sees that the model supports a two-body bound state in free space iff $a>0$~\cite{dimer_free}.

To be complete, let us recall the applicability of this narrow Feshbach resonance model.
In an experiment, the direct interaction between atoms in the open channel is not zero,
it has a finite scattering length, the so-called background scattering length
$a_{\rm bg}$ (of arbitrary sign), and a finite range, the so-called van der Waals length $b$ (of positive sign). 
If $a_{\rm bg}$ and  $b$ are  much smaller in absolute value than the scattering length $a$, the Feshbach length $R_*$, 
and the inverse of the typical relative momentum
of two particles in the gas, one can formally take the limit $a_{\rm bg}, b \to 0$ for fixed
values of $a$ and $R_*$, and one then expects to reach the present narrow Feshbach resonance model
\cite{convergence_uniforme}.

\section{Polaronic and dimeronic variational ansatz}
\label{sec:theansatz}

\subsection{Physical motivation for the ansatz}
The goal is to determine the ground state energy of a single impurity interacting with $N$ fermions.
In the two-channel model, this means that the system will in general coherently populate two sectors,
the one with $N$ fermions and the impurity, and the one with $N-1$ fermions
and one closed-channel molecule.

The exactly solvable artificial case of $\Lambda=0$ (perfectly decoupled open and closed channels)
for a finite momentum cut-off can bring some inspiration. The ground state in the $N$-fermion sector is then simply the Fermi sea
(abbreviated as ``FS")
of the $N$ fermions with the impurity in the plane wave of zero wave vector. The resulting energy
is then approximately, in the thermodynamic limit:
\be
E_{\rm FS}(N) \simeq \frac{3}{5} N E_{\rm F},
\label{eq:efs}
\ee
where we have introduced the usual Fermi energy, expressed in terms of the Fermi momentum
$k_{\rm F}$ and of the atomic density $\rho=N/V$ as
\be
E_{\rm F}=\frac{\hbar^2k_{\rm F}^2}{2m} \ \ \mbox{with}\ \ k_{\rm F}=(6\pi^2\rho)^{1/3}.
\ee
We shall use $E_{\rm FS}(N)$ as the reference energy in what follows.
The ground state in the sector with $N-1$ fermions and one closed-channel molecule is 
simply the Fermi sea of the $N-1$ fermions plus the molecule with a zero momentum center of mass,
with a resulting energy $E_{\rm FS}(N-1)+E_{\rm mol}$.
One thus has two branches of solutions, with an energy difference in the thermodynamic
limit
\be
[E_{\rm FS}(N-1)+E_{\rm mol}]-E_{\rm FS}(N) \simeq E_{\rm mol} - E_{\rm F}.
\ee
One or the other is the ground branch,
depending on the sign of $E_{\rm mol}-E_{\rm F}$, which is of course artificial.

We now turn back to the interacting case, where $\Lambda>0$ and the infinite momentum cut-off limit
is taken. As in the previous $\Lambda=0$ case,
there will be two branches, the main difference being now that the Fermi sea
is depleted by an arbitrary number of particle-hole excitations, making the problem
non-trivial. The first branch, emerging from the $N$-fermion plus impurity state,
was intensively studied in the case of zero-range single-channel models
and is called the {\sl polaronic} branch \cite{Chevy,Lobo,Svistunov,Combescot,Bruun1,Bruun2}.
The second branch, emerging from the $N-1$-fermion plus closed-channel-molecule state,
also exists in single-channel models \cite{Svistunov,Bruun1,Combescot_bs}.
It was studied in a two-channel model in \cite{Mora,Zwerger} but only in the limit of a very broad
and zero-range Feshbach resonance, where it becomes equivalent to the zero-range single-channel
case already considered in \cite{Svistunov,Bruun1,Combescot_bs}. 
In what follows, we shall call this branch the {\sl dimeronic} branch.

Following these references, we explore the two branches on a narrow Feshbach resonance
using simple variational ansatz with at most one pair of particle-hole excitations of
the Fermi sea.  To be able to determine the effective masses of the polaron and
of the dimeron, we consider the general case of a non-zero total momentum for the system: 
\be
\PP=\hbar \KK.
\ee
For the polaron we thus take the ansatz \cite{nzt}
\begin{multline}
\label{eq:ansatz_polaron}
|\psi_{\rm pol}(\PP)\rangle =  \Big(\phi\, \hat{d}_\KK^\dag
+\sum'_{\qq}  \phi_\qq \hat{b}_{\KK+\qq}^\dag \hat{u}_\qq 
\\ 
+ \sum'_{\kk,\qq}  \phi_{\kk\qq} \hat{d}^\dag_{\KK+\qq-\kk}
\hat{u}^\dag_{\kk} \hat{u}_\qq 
\Big)
|\mbox{FS}:N\rangle .
\end{multline}
The prime above the summation symbol means that the sum is restricted to $\qq$ belonging to the Fermi sea
of $N$ fermions, and to $\kk$ not belonging to that Fermi sea.
For $\KK=\mathbf{0}$, the successive terms in (\ref{eq:ansatz_polaron}) correspond in that order to the ones generated
by repeated action of the Hamiltonian $\hat{H}$ on the $\Lambda=0$ polaronic ground state.
Similarly, for the dimeron we take the ansatz \cite{nzt}
\begin{multline}
\label{eq:ansatz_dimeron}
\ket{\psi_{\rm dim}(\PP)} = \Big(\eta\, \hat{b}_\KK^\dag + \sum'_\kk \eta_\kk \hat{d}_{\KK-\kk}^\dag \hat{u}_\kk^\dag 
+ \sum'_{\kk,\qq}\eta_{\kk\qq} \hat{b}_{\KK+\qq-\kk}^\dag \hat{u}_\kk^\dag \hat{u}_\qq  \\
+ \sum'_{\kk',\kk,\qq} \eta_{\kk'\kk\qq} \hat{d}_{\KK+\qq-\kk-\kk'}^\dag \hat{u}_{\kk'}^\dag 
\hat{u}_\kk^\dag \hat{u}_\qq \Big)  \ket{\mathrm{FS}:{N-1}},
\end{multline}
where the symbols $'$ above the sums indicate that they are restricted to 
$\kk,\kk'$ out of the Fermi sea of $N-1$ fermions, and to $\qq$ inside that Fermi sea.
Here also, for $\KK=\mathbf{0}$, the order of the terms corresponds to successive actions of $\hat{H}$ on the $\Lambda=0$
dimeronic ground state.
Whereas the ansatz (\ref{eq:ansatz_dimeron}) already appeared in \cite{Mora,Zwerger} in the case $\PP=\mathbf{0}$,
the ansatz (\ref{eq:ansatz_polaron}) is to our knowledge new.

\subsection{The integral equations to be solved}
One then has to minimize the expectation value of $\hat{H}$ within each ansatz, with respect to the variational parameters
that are the complex number $\phi$ and the functions $\phi_\qq$ and $\phi_{\kk \qq}$ for (\ref{eq:ansatz_polaron}),
and that are the complex number $\eta$ and the functions $\eta_{\kk},\eta_{\kk\qq}, \eta_{\kk'\kk\qq}$
for (\ref{eq:ansatz_dimeron}) \cite{details}. As the calculations for the polaron are similar but not identical to the ones
in \cite{Chevy} we give details of the derivation in the appendix \ref{app:polaron}.
For the dimeron the calculations are quite close to \cite{Mora,Zwerger} so we directly give the resulting integral
equation in the thermodynamic limit.

For the polaronic case, one can express $\phi_{\kk\qq}$ in terms of $\phi_{\qq}$ 
and then $\phi_{\qq}$ in terms of $\phi$, 
so that one is left with a scalar implicit equation for the polaron energy counted with respect
to the $N$-fermion Fermi sea energy, 
\be
\Delta E_{\rm pol}(\PP)\equiv E_{\rm pol}(\PP) - E_{\rm FS}(N).
\label{eq:def_delta_epol}
\ee
The implicit equation in the thermodynamic limit is
\be
\label{eq:Epol}
\Delta E_{\rm pol}(\PP) =  E_\KK + \int'\!\!\frac{d^3q}{(2\pi)^3} \frac{1}{D_\qq\left[\Delta E_{\rm pol}(\PP), \PP\right]},
\ee
where the prime on the integral over $\qq$ means that it is restricted to the Fermi sea $q<k_{\rm F}$.
The function of the energy appearing in the denominator of the integrand is
\begin{multline}
\label{eq:defD}
D_\qq\left(E,\PP\right) = \frac{1}{g}- \frac{\mu k_{\rm F}}{\pi^2\hbar^2}
+ \frac{\mu^2 R_*}{\pi\hbar^4} \left(E+\varepsilon_\qq-\frac{\varepsilon_{\KK+\qq}}{1+r}\right)
\\ + \int '\!\!\frac{d^3k'}{(2\pi)^3} \left(\frac{1}{E_{\KK+\qq-\kk'}+\varepsilon_{\kk'}-\varepsilon_\qq-E}
-\frac{2\mu}{\hbar^2 k^{'2}}\right),
\end{multline}
where the prime on the integral over $\kk'$ means that it is restricted to $k'>k_{\rm F}$.
As shown in the Appendix \ref{app:integral} the integral in (\ref{eq:defD}) can be calculated 
explicitly.

For the dimeronic case, one can express $\eta, \eta_\qq$ and $\eta_{\kk\kk'\qq}$ in terms of $\eta_{\kk\qq}$.
We introduce the energy of the dimeron counted with respect to the energy of the Fermi sea of $N$ fermions: 
\be
\Delta E_{\rm dim}(\PP) \equiv E_{\rm dim}(\PP) - E_{\rm FS}(N) .
\label{eq:def_delta_edim}
\ee
In the thermodynamic limit, this energy difference is such that the following integral equation for $\eta_{\kk\qq}$
admits a non-identically-zero solution:
\be
\label{eq:system}
\int' \!\!\frac{d^3k'd^3q'}{(2\pi)^6}
\mathcal{M}[\Delta E_{\rm dim}(\PP),\PP;\kk,\qq;\kk',\qq'] \eta_{\kk'\qq'} = 0,
\ee
to be satisfied for all $\kk$ such that $k>k_{\rm F}$ and for all $\qq$ such that $q<k_{\rm F}$.
As previously mentioned, the prime on the integral sign in Eq.(\ref{eq:system}) means that
the integration over $\qq'$ is restricted to $q'<k_{\rm F}$ and the integration over $\kk'$ is
restricted to $k'>k_{\rm F}$.
The energy-dependent kernel is given by
\begin{multline}
\label{eq:kernel}
\mathcal{M}[E,\PP;\kk,\qq;\kk',\qq'] = \frac{(2\pi)^3\delta(\kk-\kk^\prime)}{F_{\kk\OO\OO}(E,\PP)} 
-\frac{(2\pi)^3\delta(\qq-\qq') }{F_{\kk'\kk\qq}(E,\PP)}
\\
- (2\pi)^6 \delta(\kk-\kk^\prime) \delta(\qq-\qq^\prime) \alpha_{\kk\qq}(E,\PP) 
+ \frac{1}{(\alpha_{\OO\OO}F_{\kk\OO\OO}F_{\kk'\OO\OO})(E,\PP)},
\end{multline}
where
\begin{equation}
\label{eq:defF}
F_{\kk^\prime\kk\qq}(E,\PP) =  E + E_{\rm F} - E_{\KK+\qq-\kk-\kk^\prime} - \varepsilon_\kk - \varepsilon_{\kk^\prime} 
+ \varepsilon_\qq,
\end{equation}
and
\be
%\begin{multline}
\label{eq:defalpha}
%\alpha_{\kk\qq}(E,\PP) = \frac{1}{g} 
%-\frac{\mu k_{\rm F}}{\pi^2\hbar^2} 
%+ \frac{\mu^2 R_*}{\pi\hbar^4}\Big(E+ E_{\rm F}+\varepsilon_\qq-\varepsilon_\kk - \frac{\varepsilon_{\KK+\qq-\kk}}{1+r}\Big) \\
%- \int' \!\!\frac{d^3k'}{(2\pi)^3}
%\Big[\frac{1}{F_{\kk^\prime\kk\qq}(E,\PP)}+\frac{2\mu}{\hbar^2k^{\prime 2}} \Big]
\alpha_{\kk\qq}(E,\PP) = D_\qq(E+E_{\rm F}-\varepsilon_\kk,\PP-\hbar\kk)
%\end{multline}
\ee
where the function $D_\qq$ defined in Eq.(\ref{eq:defD}) is calculated explicitly in
the Appendix \ref{app:integral}.
Solving the integral equation (\ref{eq:system}) for a non-identically-zero function $\eta_{\kk\qq}$
is equivalent to finding the energy $E=\Delta E_{\rm dim}(\PP)$ for a given $\PP$ such that an eigenvalue $\lambda(E,\PP)$
of the operator $M[E,\PP]$ vanishes:
\be
\label{eq:implicit_dim}
\lambda[\Delta E_{\rm dim}(\PP),\PP] =0.
\ee
Since the operator $M[E,\PP]$ is real symmetric for real $E$ and $\PP$, its eigenvalues are real,
and may be found numerically quite efficiently.
The implicit equation (\ref{eq:implicit_dim}) is then solved numerically by dichotomy or by Newton
method \cite{help} after a Gauss-Legendre discretization as in \cite{Zwerger}.

\subsection{A discrete state coupled to a continuum}
\label{subsec:adsctac}

\begin{figure}[htbp]
\begin{center}
\includegraphics[width=0.9\linewidth,clip=]{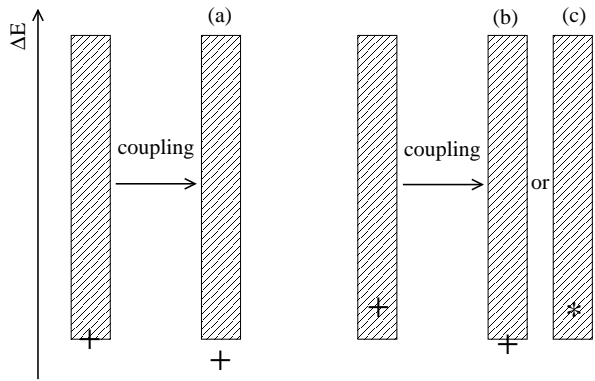}
\caption{The polaronic and dimeronic problems correspond to a discrete state coupled to a continuum. (a) When the discrete state is on the border of the
continuum it is expelled by the coupling and remains a discrete state. (b,c) When the unperturbed energy of the discrete state is inside the continuum the discrete state
is either (b) expelled or (c) washed out in the continuum where it gives rise to a resonance with a complex energy. The symbol $+$ corresponds to a discrete state while
the symbol $\ast$ corresponds to a resonance.}
\label{fig:sketch}
\end{center}
\end{figure}

Before turning to the solution of the integral equation it is useful to have a more global view of the polaronic and dimeronic problem.
In the ansatz (\ref{eq:ansatz_polaron}) and (\ref{eq:ansatz_dimeron}) it is apparent that the first contribution is a single discrete state, while the other contributions
form in the thermodynamic limit a continuum indexed by vectors $\qq,\kk$ spanning a continuous space. The problem then naturally 
falls into the well studied class of a discrete state coupled to a continuum \cite{CCT}.
There are then two cases. In the first case the discrete state is expelled from the continuum and remains a discrete state even in the presence 
of the coupling. In the second case the coupling is too weak to expel the discrete state which is then washed out in the continuum and gives
rise to a resonance with a complex energy. This resonance can still be of physical interest if the imaginary part of its energy is much smaller than
the corresponding real part. We shall encounter these two situations,  sketched in Fig.~\ref{fig:sketch}, with the polaron and the dimeron.  

In practice one can even calculate the lower border of the continuum
$\Delta E_{\rm pol}^{\rm cont}(\PP)$ 
for the polaron. One has to minimize the energy within the variational ansatz (\ref{eq:ansatz_polaron})
setting $\phi=0$. A trivial upper bound is found by setting also $\phi_\qq=0$ and, for $K<2k_{\rm F}$, we get simply $\Delta E_{\rm pol}^{\rm cont}(\PP)\le0$. However, 
including the coupling between $\phi_\qq$ and $\phi_{\kk\qq}$, we found that the actual lower border of the continuum can be negative for some values of the physical parameters 
$k_{\rm F}a$, $k_{\rm F}R_*$ and the mass ratio $r=M/m$. The reasoning is as follows: Let us assume that for a given $\qq$ the continuum eigenenergy $\Delta E_{\rm pol}^\qq$ is negative,
then it is given by the implicit equation
\cite{douvientelle}
\be
\label{eq:border_polaron}
D_\qq(\Delta E_{\rm pol}^\qq,\PP) = 0.
\ee
Since $D_\qq(\Delta E,\PP)$ is an increasing function of $\Delta E$ we can immediately distinguish two cases. In the first case, $D_\qq(0^-,\PP)$ is non positive for all $\qq$ within the 
Fermi sea. Therefore Eq.~(\ref{eq:border_polaron}) cannot have a negative solution. We conclude that the border of the continuum is simply $\Delta E_{\rm pol}^{\rm cont}(\PP)=0$.
In the second case, the lower border of the continuum $\Delta E_{\rm pol}^{\rm cont}(\PP)$ is negative, it is obtained simply by minimizing the negative solutions $\Delta E_{\rm pol}^\qq$ over $\qq$ inside the Fermi sea. 
Let us now illustrate this formal discussion with the case $\PP=\OO$. We then always found numerically that $D_\qq(\Delta E,\OO)$ for a fixed $\Delta E$ is maximal on the Fermi surface \cite{help_maximal}. 
The lower border of the continuum is then obtained by solving Eq.~(\ref{eq:border_polaron}) with $q=k_\mathrm{F}$. By expressing analytically the condition $D_{k_{\rm F}\ve_z}(0^-,\OO)>0$, we get the 
condition for this lower border to be negative:
\be
\label{eq:cfhanlbc}
\frac{1}{k_{\rm F}a} + \Big(\frac{r}{1+r}\Big)^2 k_{\rm F}R_* - \frac{1}{\pi}\Big(1-\frac{r}{1+r}\ln r\Big) > 0.
\ee

\begin{figure}[tbp]
\begin{center}
\includegraphics[width=0.9\linewidth,clip=]{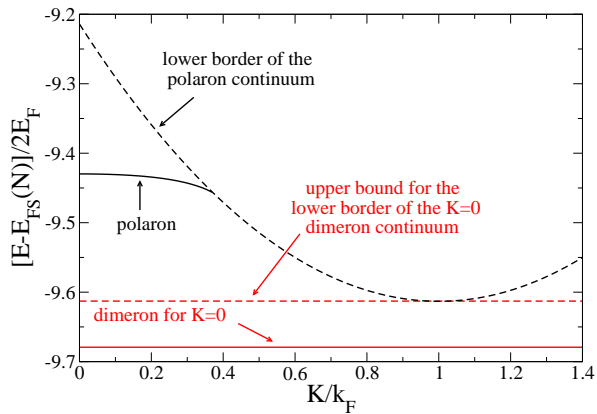}
\caption{(Color online) Polaronic energy as a function of the total momentum $\PP=\hbar\KK$ and dimeronic energy at $\PP=\OO$. The thick black line is the solution of the implicit Eq.~(\ref{eq:Epol}) over
the interval of $K$ where it admits a real solution, which corresponds to the discrete state discussed in the text. The dashed black line is the lower border of the continuum for the polaron obtained by
solving Eq.~(\ref{eq:border_polaron}) with $\qq$ on the Fermi surface~\cite{help_maximal}. The red
horizontal solid line is the solution of the implicit Eq.~(\ref{eq:implicit_dim}) for the dimeronic energy 
at $\PP=\OO$, and the red horizontal dashed line corresponds to the upper bound (\ref{eq:upper_dimeron})
for the lower border of the $\PP=\OO$ dimeron continuum. The physical parameters are: $k_{\rm F}R_*=1$, $r=0.1505$ and $1/(k_{\rm F}a)=4$.
Note that for this set of parameters the polaron has a negative effective mass at zero momentum, and the upper bound for the lower border of the $\PP=\OO$ dimeron continuum is negative.}
\label{fig:polaronic-continuum}
\end{center}
\end{figure}

We now discuss the case for the dimeron setting $\eta=0$ in Eq.~(\ref{eq:ansatz_dimeron}) and minimizing the energy. Once again, a straightforward upper bound 
for the lower border of the continuum 
$\Delta E_{\rm dim}^{\rm cont}(\PP)$ is found by setting also $\eta_\kk$ and $\eta_{\kk\qq}$ to zero and, for $K<3k_{\rm F}$, we find $\Delta E_{\rm dim}^{\rm cont}(\PP)\le0$.
A less trivial upper bound \cite{help_bound} can be obtained by taking into account the coupling of $\eta_{\kk'\kk\qq}$ to $\eta_\kk$ and $\eta_{\kk\qq}$, where we realized that the particle created by $\hat{u}_\kk^\dag$ is simply
a spectator in the energy minimization and can be factorized out. This leaves us with an ansatz to minimize which is formally equivalent to the polaronic ansatz (\ref{eq:ansatz_polaron}) with a fixed total
wavevector $\KK-\kk$ except for the replacement of $\ket{{\rm FS}:N}$ with 
$\hat{u}_\kk^\dag\ket{{\rm FS}:N-1}$. To obtain this upper bound it remains
to minimize over $\kk$ out of the Fermi sea
\be
\label{eq:border_dimeron}
\Delta E_{\rm dim}^{\rm cont}(\PP) \le \min_{\kk,k>k_{\rm F}} \Big[ \Delta E_{\rm pol}^{\rm gs}(\PP-\hbar\kk)+\frac{\hbar^2k^2}{2m} -E_{\rm F} \Big],
\ee
where $\Delta E_{\rm pol}^{\rm gs}(\PP)$ is the {\it absolute} ground state of the polaronic ansatz (\ref{eq:ansatz_polaron}) (that is taken over the discrete and continuous spectra).
We now discuss this result for vanishing total momentum $\PP=\OO$. In this situation we observed numerically that the global minimum of Eq.~(\ref{eq:border_dimeron}) is obtained
at the Fermi surface so that
\be
\label{eq:upper_dimeron}
\Delta E_{\rm dim}^{\rm cont}(\OO) \le \Delta E_{\rm pol}^{\rm gs}(\hbar k_{\rm F}\ve_z),
\ee
and can be negative depending on the physical parameters $k_{\rm F}a$, $k_{\rm F}R_*$ and the mass ratio $r=M/m$. Fig.~\ref{fig:polaronic-continuum} summarizes the discussion above, where for a given choice of the physical parameters we plot the lower border of the continuum for the polaron
as well as the upper bound for the lower border of the $\PP=\OO$ dimeron continuum. 

To conclude we note that the qualitative features that we have discussed apply only for the specific variational ansatz (\ref{eq:ansatz_polaron}) and (\ref{eq:ansatz_dimeron}) that we have considered. In an exact theory
with an arbitrary number of particle-hole excitations the polaronic and dimeronic solutions at $\PP\ne\OO$ are actually resonances rather than stationary states because they can radiate particle-hole pairs of
arbitrary small energies~\cite{Stringari}. This implies that the gap between the discrete state and the continuum sketched in Fig.~\ref{fig:sketch} will close.

\section{Properties of the two branches at $\PP=\OO$}
\label{sec:properties}
\begin{figure*}[htbp]
\begin{center}
\includegraphics[width=1\linewidth,clip=]{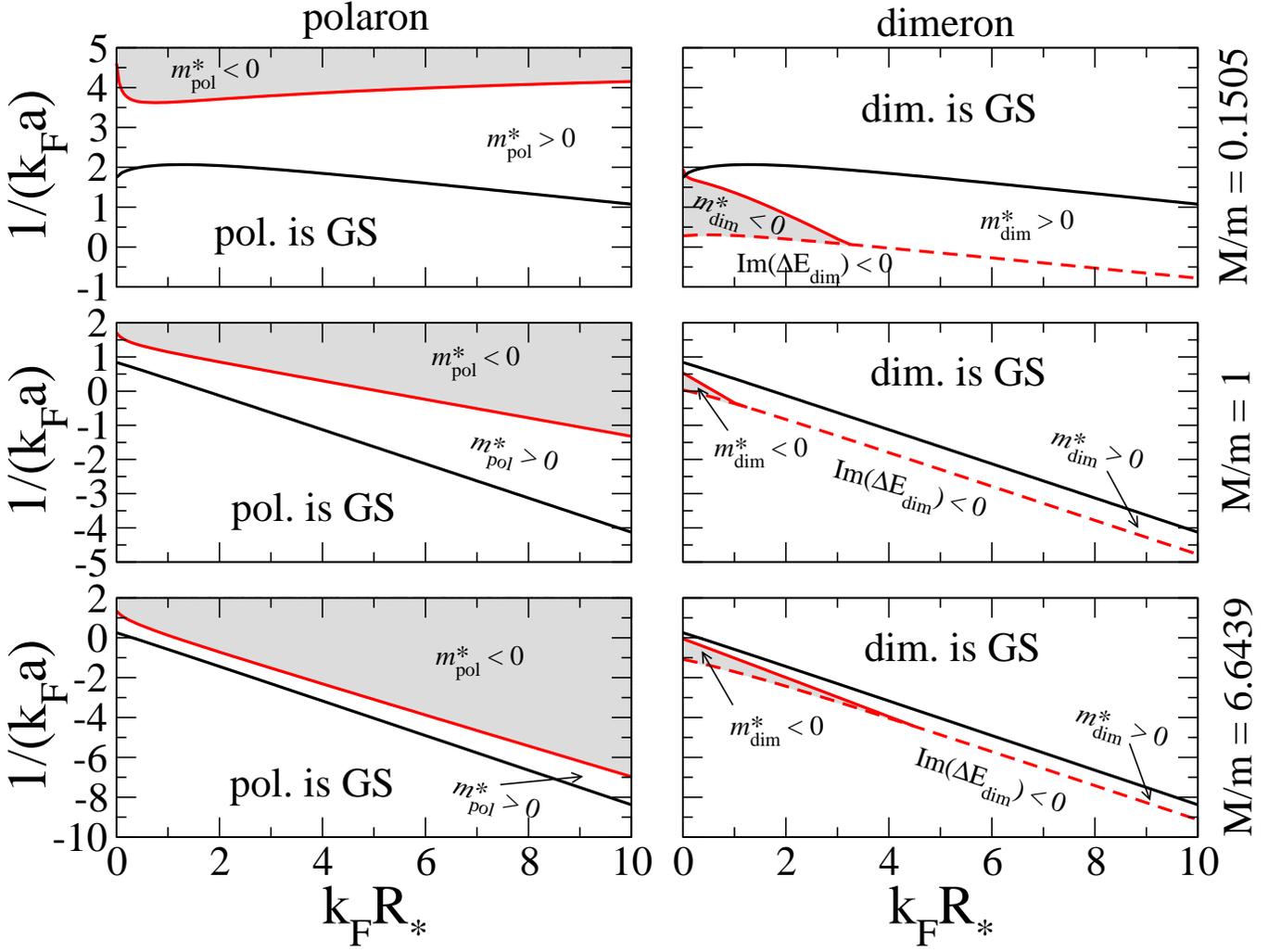}
\caption{(Color online) Phase diagram of the system at $\PP=\OO$ and zero total angular momentum in the $(k_{\rm F}R_*)$--$(1/k_{\rm F}a)$ plane. The left panels describe the polaronic sector while the right
panels belong to the dimeronic sector. GS indicates that the corresponding branch is the ground state,
and the black solid line marks the polaron-dimeron transition.  $m^*_{\rm pol(dim)}>0$
indicates that the corresponding branch is metastable with a positive effective mass, the gray-shaded area 
is the region where the
corresponding branch is thermodynamically unstable due to a negative effective mass $m^*_{\rm pol(dim)}<0$,
the red solid line indicates where the inverse effective mass is zero. 
For the dimeron, $\mathrm{Im}(\Delta E_{\rm dim})<0$
indicates that its energy is complex with negative imaginary part 
(which implies a positive real part, so that the dashed red lines correspond to $\Delta E_{\rm dim}=0$). 
Note that in the ground state of our variational ansatz the effective mass is always positive, as expected, except in a narrow region 
in the upper-right panel (see subsection~\ref{sec:strange}).}
\label{fig:phase-diagram}
\end{center}
\end{figure*}
We calculate the energies of the two branches by solving Eq.~(\ref{eq:Epol}) for the polaron and Eq.~(\ref{eq:implicit_dim}) for the dimeron, at $\PP=\OO$.
In both cases we assume spherical symmetry in the system. This amounts to assuming that the function $\eta_{\kk\qq}$ appearing in the integral equation 
(\ref {eq:system}) has total angular momentum $\ell=0$, so that we can write  
\be
\eta_{\kk,\qq} = f(k,q,\theta),
\ee
with $\theta$ being the angle between the vectors $\kk$ and $\qq$. For the polaronic branch, assuming spherical symmetry means that the function 
$\phi_\qq$ appearing in the ansatz (\ref{eq:ansatz_polaron}) is spherically symmetric, as explained in the Appendix \ref{app:polaron}. 

In Fig.~\ref{fig:phase-diagram} we plot the phase diagram of the system in the $(k_{\rm F}R_*)$--$(1/k_{\rm F}a)$ plane for three different values of
the mass ratio:  From top to bottom respectively $r=0.1505$ corresponding to a $^6$Li impurity into a Fermi sea of $^{40}$K, $r=1$, and $r=6.6439$, which corresponds to 
a $^{40}$K impurity into a Fermi sea of $^6$Li. In the polaronic sector (left panels) we identify three
different regions: (i) the region below the black solid line is where the polaron is the ground branch, (ii) the region between the black solid line and the 
lower boundary of the gray-shaded area is where the polaron is metastable having a positive effective mass $m^*_{\rm pol}>0$, (iii) the region delimited
by the gray-shaded area is where the polaron is thermodynamically unstable due the the fact that its effective mass is negative $m^*_{\rm pol}<0$.
Instead, in the dimeronic branch (right panels) we distinguish between four different regions: (i) the region above the black solid line is where the dimeron is the ground branch,
(ii) the region between the black solid line and the red dashed line (excluding the gray-shaded area) is where the dimeron is metastable featuring  a
positive effective mass $m^*_{\rm dim}>0$, (iii) the gray-shaded area is where the dimeronic effective mass is negative $m^*_{\rm dim}<0$ and indicates a thermodynamic instability,
(iv) the region below the red dashed line is where the dimeronic branch becomes complex, with an energy 
$\Delta E_{\rm dim}$ having a positive real part and a negative imaginary part.
In what follows we derive and discuss the different features of the phase diagram shown in Fig.~\ref{fig:phase-diagram}.
The main effect of having a narrow Feshbach resonance in the regime $k_{\rm F}R_* \gg 1$ is that the transition from having the polaron as the ground state to having the dimeron as the ground state 
takes place for negative values of the scattering length, whereas it takes place for positive scattering lengths in the zero-range limit $k_{\rm F}R_*\to0$.

\subsection{Polaron to dimeron transition}
\begin{figure}[tbp]
\begin{center}
\includegraphics[width=1.0\linewidth,clip=]{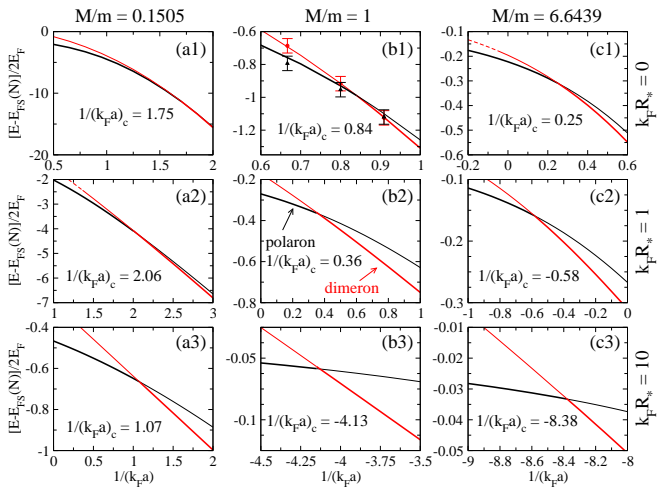}
\caption{(Color online) Crossing between the polaronic energy Eq. (\ref{eq:Epol}) and dimeronic energy Eq. (\ref{eq:implicit_dim}), for different values of the mass
ratio: $r=0.1505$ (column a), $r=1$ (column b), $r=6.6439$ (column c), and of the Feshbach length, respectively $k_{\rm F}R_*=0$ (row 1), $k_{\rm F}R_*=1$ (row 2) and $k_{\rm F}R_*=10$ (row 3).
Thick lines correspond to the ground state branch while the thin lines indicate 
that the corresponding branch is metastable. Instead the dashed lines correspond to a negative value of the effective mass.
The red circles (dimeron) and black triangles (polaron) with error bars 
are the Monte Carlo results of Ref. \cite{Svistunov}.}
\label{fig:crossings}
\end{center}
\end{figure}
\begin{figure}[htbp]
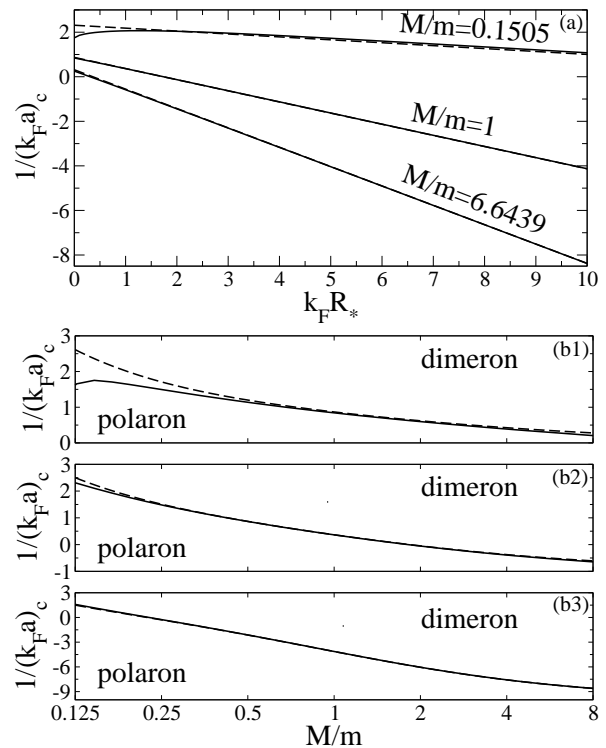

\begin{center}
\includegraphics[width=0.9\linewidth,clip=,scale=1.5]{fig5a.eps}
\includegraphics[width=0.9\linewidth,clip=]{fig5b.eps}
\caption{Dimeron-polaron crossing point $1/(k_{\rm F}a)_c$ as a function of various parameters. (a) As a function of the Feshbach length $R_*$, 
for three values of the mass ratio ($r=0.1505$, $r=1$, $r=6.6439$ from top to bottom).
(b) As a function of the mass ratio $r=M/m$ for three values of the Feshbach length: (b1) $k_{\rm F}R_*=0$,
(b2) $k_{\rm F}R_*=1$, (b3) $k_{\rm F}R_*=10$.
For each of the figures, $1/(k_{\rm F} a)_c$ is the line in the plane $(k_{\rm F} R_*)$--$(1/k_{\rm F} a)$ 
or $(M/m)$--$(1/k_{\rm F} a)$
that separates the region where the dimeron is the ground state (above that line) from the region where
the polaron is the ground state (below that line). Solid lines: Numerical solution from the full polaronic and dimeronic
ansatz. Dashed lines: Asymptotic analytical prediction for $k_{\rm F} R_*\to +\infty$ (\ref{eq:kfac_analy}). 
The dashed line is barely distinguishable from the solid line for $r=1$ in (a) and for (b3).
 Although it does not {\sl a priori} make sense, we have plotted this asymptotic prediction
even in the regimes $k_{\rm F} R_* \lesssim 1$ or  $-1 \lesssim 1/(k_{\rm F} a)_c$ to show that it remains quite close to the numerical results
even in these non-asymptotic regimes.
}
\label{fig:crossing-3}
\end{center}
\end{figure}
As discussed in Ref.~\cite{Svistunov} using diagrammatic Monte Carlo simulations, for a unity impurity-to-fermion mass ratio ($r=1$) and for a wide Feshbach resonance ($k_\mathrm{F}R_* = 0$) 
the polaron to dimeron transition is characterized by a crossing between the two branches, the crossing point being located at $1/(k_\mathrm{F}a)_c \simeq 0.90$. This is shown in Fig.~\ref{fig:crossings}~(b1) where we plot the energies of the two branches calculated with Eq.~(\ref{eq:Epol}) and
(\ref{eq:implicit_dim}) at $\PP=\OO$, together with the Monte Carlo results of Ref.~\cite{Svistunov}. On the right hand side of the crossing point $1/(k_\mathrm{F}a) > 1/(k_\mathrm{F}a)_c$ the ground branch is dimeronic, while  
on the left hand side of the crossing point $1/(k_\mathrm{F}a) < 1/(k_\mathrm{F}a)_c$ the ground state is polaronic.
%The agreement between the latter and our variational calculations is remarkable and we locate the crossing point at $1/(k_\mathrm{F}a)_c \simeq 0.84$, which is approximately $6 \%$ away from the Monte Carlo value. 
In Fig.~\ref{fig:crossings}~(b2) and (b3) we show how the energies of the polaron and of the dimeron are modified by a narrow resonance. 
In Fig.~\ref{fig:crossings}~(a) we plot the polaronic and dimeronic energies for the mass ratio $r=0.1505$, which corresponds to a $^6$Li impurity into a Fermi sea of $^{40}$K, while
in Fig.~\ref{fig:crossings}~(c) we plot the polaronic and dimeronic energies for the mass ratio $r=6.6439$, corresponding to a $^{40}$K impurity into a Fermi sea of $^6$Li.
%
%, respectively at $k_\mathrm{F}R_* = 1$ we locate
%it at $1/(k_\mathrm{F}a)_c \simeq 0.36$, while at $k_\mathrm{F}R_* = 10$ the crossing point surprisingly shifts on the BCS side $1/(k_\mathrm{F}a)_c \simeq-4.13$. 
%For the mass ratio $r=6.6439$, corresponding to a $^{40}$K impurity into a Fermi sea of $^6$Li, we find again that the crossing point for a broad resonance  
%$1/(k_\mathrm{F}a)_c\simeq0.25$ shifts progressively towards the BCS
%region as the parameter $k_\mathrm{F}R_*$ becomes larger as shown in the (c) panels of Fig.~\ref{fig:crossings}. At $k_\mathrm{F}R_*=1$ we find the crossing point 
%to be $1/(k_\mathrm{F}a)_c \simeq-0.58$, and at $k_\mathrm{F}R_*=10$ we find it to be as low as $1/(k_\mathrm{F}a)_c \simeq-8.38$. Instead for the mass ratio $r=0.1505$, which corresponds to a $^6$Li impurity into a Fermi sea of $^{40}$K the situation is different as shown in the (a) panels of Fig.~\ref{fig:crossings}: By changing the Feshbach length from $k_\mathrm{F}R_*=0$ to $k_\mathrm{F}R_*=1$ the
%crossing point undergoes a shift towards the BEC region from  $1/(k_\mathrm{F}a)_c \simeq1.75$ to $1/(k_\mathrm{F}a)_c \simeq 2.06$, while
%at $k_\mathrm{F}R_*=10$ the crossing point moves back towards the BCS region but remains positive $1/(k_\mathrm{F}a)_c \simeq1.08$. 
%
To summarize, in Fig.~\ref{fig:crossing-3} (a) we plot the polaron-to-dimeron crossing point for the above mass ratios and for $0 \le k_\mathrm{F}R_* \le 10$. 
For a broad resonance $k_\mathrm{F}R_*=0$~(a) we find that the crossing point is positive for the whole range of plotted mass ratios $r$. 
At $r=0.1505$, for the whole range of Feshbach lengths we find that $1/(k_\mathrm{F}a)_c$ is positive on the range covered by the figure, but for larger $k_\mathrm{F}R_*$ (not shown) 
we find that the crossing point eventually becomes negative.
For $r=1$ and $r=6.6439$ the crossing point is a monotonically decreasing function of $k_\mathrm{F}R_*$ with a faster decrease for a higher 
mass ratio $r$.
For completeness, in Fig.~\ref{fig:crossing-3} (b) we plot the crossing point as a function of the mass ratio $(r=M/m)$, for
different values of the Feshbach length.

Finally, let us discuss the validity of our variational approach
%. We have already stated that for a broad resonance and $r=1$ both the polaronic and dimeronic 
%ansatz at $\PP=\OO$ show a remarkable agreement with the corresponding energies calculated with Monte Carlo simulations in Ref.~\cite{Svistunov}, and as can
%be seen in Fig.~\ref{fig:crossings}~(b1) the variational energies lay within the Monte Carlo error bars.  
%However, at present there are no Monte Carlo data available for narrow resonances. We can nevertheless ask ourselves what is the error introduced by
where we cut the Hilbert space at one pair of particle-hole excitations both for the polaronic and for the dimeronic ansatz.
For the polaronic ansatz, it was shown in the second paper of Ref.~\cite{Combescot} for a broad resonance and $r=1$ that going from one pair of particle-hole excitations to two pairs
brings typically a $10^{-2}$ correction to the energy in units of $E_{\rm F}$.
For the dimeronic ansatz one may consider the zeroth order approximation, in which no particle-hole excitations are present, and calculate the correction induced by the
first order approximation Eq.~(\ref{eq:ansatz_dimeron}) in which one pair of particle-hole excitations is present. Setting $\eta_{\kk\qq}=\eta_{\kk'\kk\qq}=0$
into Eq.~(\ref{eq:ansatz_dimeron}) leads to a Cooper-type ansatz given by the following expression
\begin{multline}
\label{eq:ansatz_Cooper}
\ket{\psi_{\rm Cooper}(\PP)} = \bigg(\eta\, \hat{b}_\KK^\dag + \sum'_\kk \eta_\kk \hat{d}_{\KK-\kk}^\dag \hat{u}_\kk^\dag \bigg)  \ket{\mathrm{FS}:{N-1}},
\end{multline}
which, after minimization of the expectation value of $\hat{H}$, at $\PP=\OO$ leads to the Cooper's energy  $\Delta E_\mathrm{Cooper}(\OO) = E_\mathrm{Cooper}-E_\mathrm{FS}(N)$ given by the implicit equation
\be
\label{eq:Cooper_def}
\alpha_{\OO\OO}[\Delta E_\mathrm{Cooper}(\OO),\OO]\equiv D_{\OO}(\Delta E_\mathrm{Cooper}(\OO)+E_{\rm F},\OO) = 0,
\ee 
where the definition of $\alpha$ in Eq.~(\ref{eq:defalpha}) was used. We can thus estimate the accuracy of the dimeronic ansatz (\ref{eq:ansatz_dimeron}) by calculating the relative energy difference between the Cooper and the dimeronic ansatz at the energy crossing point
between the polaron and the dimeron:
\be
\delta = \left(\frac{|\Delta E_\mathrm{dim}(\OO)-\Delta E_\mathrm{Cooper}(\OO)|}{|\Delta E_\mathrm{dim}(\OO)|}\right)
_{k_{\rm F} a=(k_{\rm F} a)_c}.
\ee
For a Feshbach length $k_{\rm F} R_*=10$, and for a mass ratio $r=M/m$ ranging from $1/8$ to $8$,
we have found a maximal value for $\delta$ equal to $10^{-1}$,
indicating that we have already reached a
good convergence of the dimeronic energy $\Delta E_\mathrm{dim}(\OO)$ in terms of the number of particle-hole excitations.

\subsection{Effective mass}
\label{subsec:effective-mass}
\begin{figure}[htbp]
\begin{center}
\includegraphics[width=1.0\linewidth,clip=]{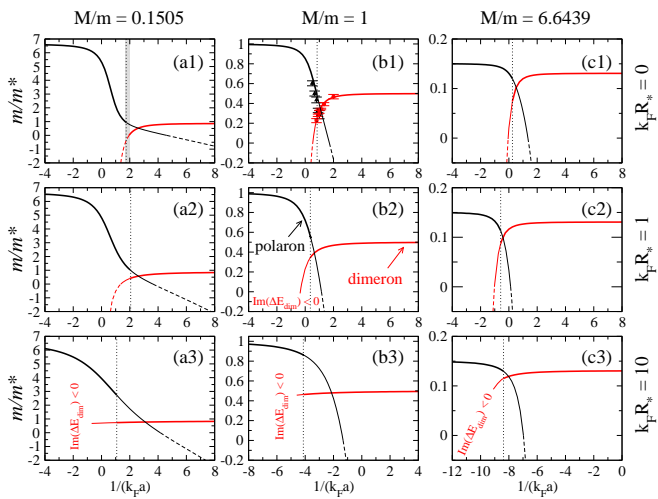}
\caption{(Color online) Inverse effective mass of the dimeron (red lines) and of the polaron (black lines), for different values of the 
Feshbach length, respectively $k_{\rm F}R_*=0$ (row 1), $k_{\rm F}R_*=1$ (row 2), $k_{\rm F}R_*=10$ (row 3), and different values of the mass ratio: 
$r=0.1505$ (column a), $r=1$ (column b) and $r=6.6439$ (column c).
The vertical dotted lines are in correspondence of the 
crossing value $1/(k_{\rm F}a)_c$: On the left of this line, the ground state is polaronic while on the right the system
has a dimeronic ground state. Thick lines correspond to the ground state, thin lines indicate where the system is metastable and dashed lines where the 
effective mass is negative. The red circles (dimeron) and black triangles (polaron) with error bars
correspond to the Monte Carlo results of 
Ref.~\cite{Svistunov}.
The inverse effective mass of the polaron tends to $m/m_{\rm pol}^* \rightarrow 1/r$ in the limit where $a\rightarrow 0^-$, while the inverse effective mass of the dimeron tends
to $m/m_{\rm dim}^* \rightarrow 1/(1+r)$ in the limit where $a\rightarrow 0^+$. The gray-shaded area in (a1) indicates an instability of the system 
(within our $\PP=\mathbf{0}$, $\ell=0$ variational calculation): The ground branch is dimeronic 
and its effective mass is negative.}
\label{fig:meff}
\end{center}
\end{figure}

The polaron and dimeron effective masses $m^*$ are 
defined by the expansions for a vanishing total momentum~$\PP$:
\be
\label{eq:quadratic}
\Delta E(\PP) = \Delta E(\OO) + \frac{P^2}{2 m^*} + O (P^4).
\ee
The calculation technique for these effective mass are 
presented in detail in the Appendix \ref{app:dim-meff}. 
A negative effective mass indicates in our model a thermodynamic instability:
It cannot manifest itself for an isolated evolution 
since our Hamiltonian conserves the total momentum, but it will manifest
itself if the system is weakly coupled to a thermal bath, or in 
a harmonically trapped system, as in real experiments, since $\PP$ 
is then not anymore a constant of motion and will slowly drift
away from $\OO$ to reduce the system energy.

In Fig.~\ref{fig:meff} we plot both the dimeronic and the polaronic inverse effective mass as a function of $1/(k_{\rm F}a)$ for different values of the Feshbach length
and different values of the mass ratio. Notice that in panel (b1), corresponding to a broad resonance and $r=1$, we also plot the Monte Carlo data
from Ref.~\cite{Svistunov}; while the dimeronic inverse effective mass shows a very good agreement with the Monte Carlo data, the polaronic inverse effective mass 
is slightly larger than the corresponding Monte Carlo value. As explained in the second paper of Ref.~\cite{Combescot} this difference is due to the truncation of the Hilbert space 
spanned by the polaronic ansatz to at most one pair of particle-hole excitations. In the same reference it was shown that including two pairs of particle-hole excitations 
in the polaronic ansatz leads to an excellent agreement with the Monte Carlo data. However, the inclusion of two particle-hole excitations is beyond the scope
of this work.

For all mass ratios considered here, in the limit where $a\rightarrow 0^+$ the effective mass of the dimeron tends to the sum of the mass of the impurity plus the mass of a fermion $m_{\rm dim}^*=m+M$. 
On the polaronic branch, in the limit where $a\rightarrow 0^-$ the polaron effective mass tends
to the mass of the impurity $m_{\rm pol}^*=M$, since as $a \to 0^-$ the system tends to a configuration of an ideal Fermi gas of $N+1$ non-interacting particles.
In between these two limits, depending on the mass ratio and on the Feshbach length we find different values of the effective masses as shown in Fig.~\ref{fig:meff}.
The vertical dotted lines indicate where the polaron to dimeron crossing point takes place; for $1/(k_\mathrm{F}a)$ smaller than this value the ground state of
the system is polaronic, while for $1/(k_\mathrm{F}a)$ larger than this value the ground state is dimeronic.
Thick lines indicate that the corresponding branch is the ground state while thin lines represent a metastable branch. Instead, dashed lines are in correspondence 
of  negative values of the effective mass indicating an instability of the corresponding branch. Notice that for the panels labeled as (3), and for
the panel (b2), the curve of the dimeronic effective mass has an abrupt stop at a negative value of $1/(k_{\rm F}a)$. This point is where the dimeronic branch becomes complex
with a negative imaginary part, as explained in subsection~\ref{subsec:continuation}.

We find that both for the polaron and for the dimeron there exist divergent values of the effective mass. For the polaronic branch, $m/m_{\rm pol}^* = 0$ in the region 
where the polaron is not any more the ground state (on the right hand side of the vertical dotted line). 
Similarly, the dimeronic effective mass diverges $m/m_{\rm dim}^* = 0$ in the region where the dimeron is not any more the ground state (on the left hand side of the vertical dotted line).
However, we observe that for the panel (a1) corresponding to $k_\mathrm{F}R_*=0$, there exists a point $1/(k_\mathrm{F}a)$ in the region where the dimeron has a lower energy than the polaron where the dimeron effective mass diverges,
and in the region between the divergent point ($m/m_{\rm dim}^*=0$) and the polaron to dimeron crossing point, the dimeron effective mass becomes negative.
We call this region the instability region, indicated by the gray shaded area in the panel (a1). The question on the nature of the true ground state of the system
in this instability region will be addressed in Sec.~\ref{sec:strange}.
We note however that by increasing the Feshbach length to the value $k_\mathrm{F}R_*=1$ the instability region disappears as can be seen in Fig.~\ref{fig:meff}~(a2).

\subsection{Closed channel population}
\label{sub:closed}
\begin{figure}[tbp]
\begin{center}
\includegraphics[width=1.0\linewidth,clip=]{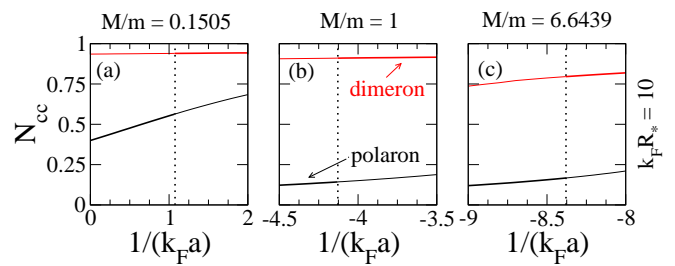}
\caption{(Color online) Closed channel population (with $\PP=\OO$) 
as a function of $1/(k_{\rm F}a)$ for different values of the mass ratio: $r=0.1505$ (a), $r=1$ (b) and $r=6.6439$ (c). These correspond to the three bottom panels of Fig \ref{fig:crossings}, where $k_{\rm F}R_* = 10$.
Thick lines correspond to the ground state while thin lines indicate that the corresponding branch is metastable. The vertical dotted lines are in correspondence of the polaron-to-dimeron 
crossing point: On the left the ground state is polaronic and on the right it is dimeronic.}
\label{fig:NccR10}
\end{center}
\end{figure}
An interesting quantity to consider is the mean number of closed channel molecules, defined as
\be
N_\mathrm{cc} =  \sum_{\kk} \langle \hat{b}_\kk^\dagger \hat{b}_\kk \rangle,
\ee
where the expectation value is taken in the quantum state of the system. This
quantity was recently measured in the BEC-BCS crossover for the spin balanced Fermi gas
by photoassociation techniques \cite{Hulet_mol}. In our single impurity problem,
$N_\mathrm{cc}$ is less than or equal to one, and can also  be interpreted 
as the occupation probability (or population) of the closed channel.

When the system is prepared in an eigenstate of energy $E$, and relative energy $\Delta E= E-E_\mathrm{FS}(N)$
with respect to the non-interacting Fermi sea, with a fixed total momentum $\PP$ if desired,
a direct application of the Hellmann-Feynman theorem
gives \cite{Tarruell}:
\be
N_{\rm cc} = \frac{\partial \Delta E}{\partial E_{\rm mol}},
\ee
the derivative being taken for fixed values of the other parameters of the Hamiltonian. Taking as a parameter
the inverse scattering length rather than $E_{\rm mol}$, as in \cite{Tarruell}, and using Eqs.(\ref{eq:Emol}) and (\ref{eq:Rstar}) 
we can rewrite $N_\mathrm{cc}$ in the more convenient form
\be
\label{eq:Ncc_def}
N_\mathrm{cc} =- \frac{\partial \Delta E}{\partial (1/a)} \frac{2\mu R_*}{\hbar^2}.
\ee
This shows that $N_{\rm cc}$ vanishes in the broad resonance limit $R_*\to 0$,
where $N_{\rm cc}/R_*$ is proportional 
to the so-called ``contact" \cite{Tan,Braaten_contact} first introduced in one-dimensional systems
in \cite{Lieb,Olshanii}. Note that, within our variational ansatz,
one has also $N_\mathrm{cc} = | \eta |^2 +  \sum'_{\kk,\qq} | \eta_{\kk\qq} |^2$ for the dimeron and $N_\mathrm{cc} = \sum'_{\qq} | \phi_{\qq} |^2$ for the polaron, which are less convenient to evaluate
than (\ref{eq:Ncc_def}).

A straightforward calculation is to find $N_\mathrm{cc}$, both for the polaron and for the dimeron, in the limit 
$a\to0^+$ but keeping $R_*$ fixed. In this limit both the polaronic and the dimeronic energies are asymptotically
equivalent to the energy of a dimer in free space $E_\mathrm{d}=-\hbar^2q_\mathrm{rel}^2/2\mu$, with $q_\mathrm{rel}$ given in \cite{dimer_free}.
One can also easily check that $q_\mathrm{rel} \sim 1/\sqrt{R_*a}$, so that
%\be
%-\frac{\partial E_\mathrm{d}}{\partial (1/a)} = \frac{\hbar^2q_\mathrm{rel}}{\mu\sqrt{1+4R_*/a}}, 
%\ee
%which, for a fixed $R_*$, has the simple limit
%\be
%\lim_{a \to 0^+} \frac{\hbar^2q_\mathrm{rel}}{\mu\sqrt{1+4R_*/a}} = \frac{\hbar^2}{2\mu R_*}.
%\ee
in this simple limit the closed channel population becomes unity: 
\be
\lim_{a \to 0^+} N_\mathrm{cc} = 1,
\ee
independently of the mass ratio as long as the Feshbach length is finite. Therefore, starting from this limit by increasing the value of the scattering length
$N_\mathrm{cc}$ must decrease and one has to numerically calculate its value using Eq.~(\ref{eq:Ncc_def}). In Fig.~\ref{fig:NccR10} we plot the closed
channel molecule population both for the polaron and for the dimeron in correspondence of the polaron to dimeron crossing at $k_{\rm F}R_* = 10$ and for different values of the mass ratio. The three panels (a), (b) 
and (c) correspond to the three lower panels of Fig.~\ref{fig:crossings}. Notice that for the three cases considered here the polaronic closed channel population is always smaller than the corresponding dimeronic
closed channel population. 
Furthermore, as we will derive analytically in Sec.~\ref{sec:ariantwalapz}, the dimeronic (resp.\ polaronic) closed channel population tends to unity (resp.\ zero) at the crossing
for $k_{\rm F} R_*\to +\infty$.

\subsection{Right at resonance}
\begin{figure}[tbp]
\begin{center}
\includegraphics[width=0.9\linewidth,clip=]{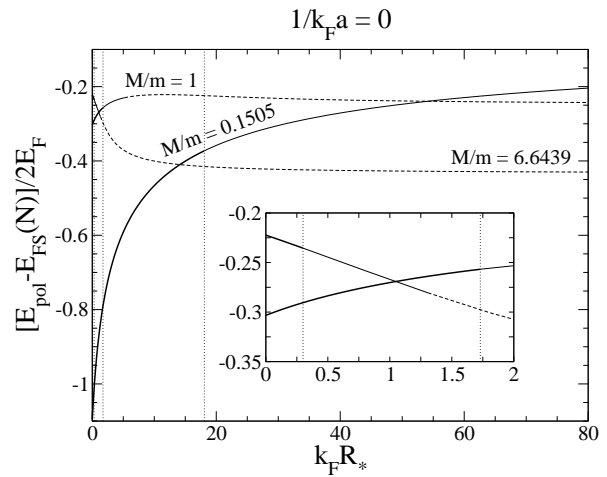}
\caption{Energy of the polaron calculated at $1/(k_{\rm F}a) = 0$, as a function of $k_{\rm F}R_*$ for different values of the mass ratio: 
$r=1$, $r=0.1505$, and $r=6.6439$. Thick lines indicate a polaronic ground state, while thin lines indicate that the polaronic branch is metastable and the ground state is dimeronic. 
Dashed lines indicate that the effective mass is negative. 
Vertical dotted lines indicate the crossing point between the dimeronic and polaronic branches
(from left to right: $r$ in decreasing order).
The inset is a magnification.
}
\label{fig:polaron-inva0}
\end{center}
\end{figure}
\begin{figure}[tbp]
\begin{center}
\includegraphics[width=0.9\linewidth,clip=]{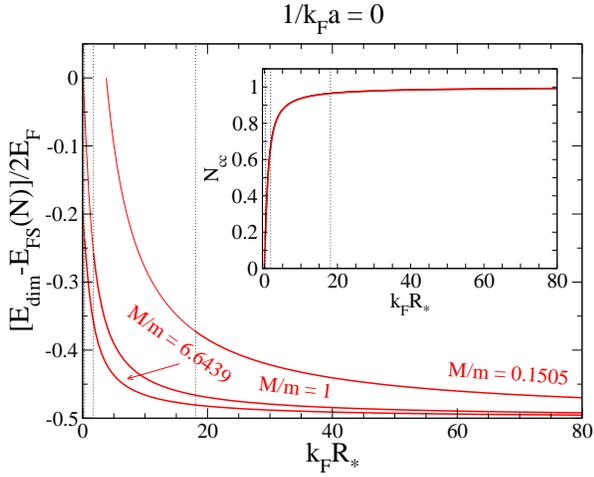}
\caption{(Color online) Energy of the dimeron calculated at $1/(k_{\rm F}a) = 0$, as a function of $k_{\rm F}R_*$ for different values of the mass ratio:
$r=1$, $r=0.1505$, and $r=6.6439$. Thick lines indicate a dimeronic ground state, while thin lines indicate that the dimeronic branch is metastable and the ground state is polaronic. 
An abrupt stop of the energy lines indicate that the dimeronic branch becomes complex.
Vertical dotted lines indicate the crossing point between the dimeronic and polaronic branches
(from left to right: $r$ in decreasing order).
The inset shows the corresponding closed channel occupations for the three different mass ratios;
the curves overlap and cannot be distinguished at the scale of the figure.}
\label{fig:dimeron-inva0}
\end{center}
\end{figure}

An interesting point in the phase diagram is to look what happens at resonance, i.e at $1/(k_{\rm F}a) = 0$. In Fig.~\ref{fig:polaron-inva0} and Fig.~\ref{fig:dimeron-inva0}
we plot respectively the energy of the polaron and the energy of the dimeron as a function of $k_{\rm F}R_*$ for different values of the mass ratio.
The thick lines indicate where the polaronic (resp.\ dimeronic) branch is the ground state. Instead the thin lines indicate that the corresponding branch is metastable. 
At $k_{\rm F}R_*=0$ the ground state belongs to the polaronic branch for the three mass ratio considered here. 
When one increases $k_{\rm F}R_*$ the ground state eventually becomes
dimeronic, at a value of $k_{\rm F}R_*$ depending on the mass ratios. For $r=1$, the transition occurs at 
$k_{\rm F}R_*\simeq 1.74$; for $r=6.6439$ it occurs at $k_{\rm F}R_*\simeq 0.31$ (see inset of Fig.~\ref{fig:polaron-inva0}); for $r=0.1505$, it occurs at the large value $k_{\rm F}R_*\simeq18.09$.
As can be seen from the inset of Fig.~\ref{fig:dimeron-inva0}, for $R_*\to \infty$,
the closed channel population saturates to unity. Also,
the dimeronic energy $\Delta E_{\rm dim}(\OO)$ tends to $-E_\mathrm{F}$,
as readily obtained within the Cooper approximation (\ref{eq:Cooper_def})
\cite{limmef}.

\subsection{Dimeronic branch: Analytical continuation}
\label{subsec:continuation}
\begin{figure}[tbp]
\begin{center}
\includegraphics[width=1.0\linewidth,clip=]{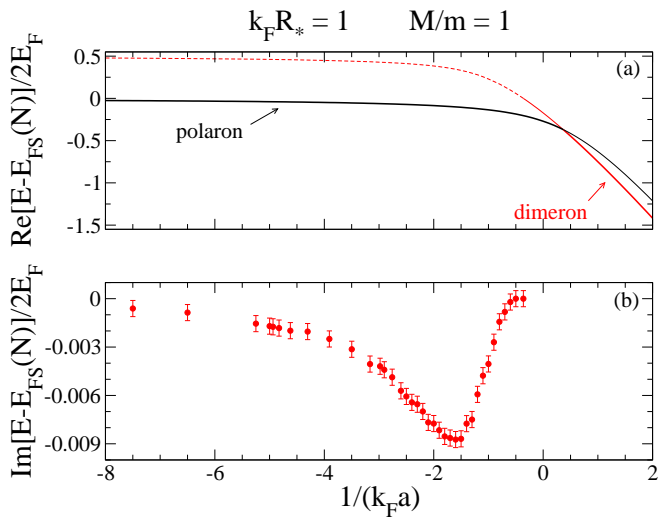}
\caption{(Color online) Energy of the dimeronic branch $\Delta E_{\rm dim}(\OO) = E_{\rm dim}(\OO) - E_{\rm FS}(N)$ at $\PP=\OO$, calculated 
for equal masses $r=1$ and at $k_{\rm F}R_* = 1$ (in red): (a) real part, (b) imaginary part (with error bars
indicating the numerical uncertainty due to discretization of the momentum variables).
The real negative branch for $\Delta E_{\rm dim}(\OO)$ is either below (red thick line)
or above (red thin line) the polaron energy $\Delta E_{\rm pol}(\OO)$ (black line). 
As soon as $\Delta E_{\rm dim}(\OO)$ crosses the zero threshold 
the dimeronic energy becomes complex with a small imaginary part that tends to $0^-$ when $a\rightarrow 0^-$; 
in (a) the real part of the dimeronic energy is then indicated with a red dashed line.}
\label{fig:complex-ener}
\end{center}
\end{figure}

To calculate the energy of the dimeronic branch at zero momentum $\PP=\OO$, one has to evaluate the integral in Eq.~(\ref{eq:defalpha}) for $E=\Delta E_\mathrm{dim}(\OO)$.  As explained in the Appendix~\ref{app:integral} this integral is well defined as long as $E<0$; even in the worst case $E=0$ it is found
to be convergent.
However, we always found that the relative dimeronic energy $\Delta E_\mathrm{dim}(\OO)$ calculated with Eq.~(\ref{eq:implicit_dim}) reaches $E=0$ at some value $1/(k_\mathrm{F}a)_x$ with a finite derivative $\partial_aE > 0$ in that point. This is shown for example in Fig.~\ref{fig:complex-ener}~(a) with the full line vanishing at $1/(k_\mathrm{F}a)_x \simeq -0.35$.
It is therefore natural to ask if it is possible to analytically continue this curve and allow for complex values of the dimeronic energy:
\be
\Delta E_{\rm dim}(\OO) = E_0 - i\frac{\hbar\Gamma}{2}, \;\;\; \mathrm{where} \;\;\; E_0,\Gamma > 0,
\ee
with $\Gamma$ being the decay rate. Such complex energy solutions do not correspond anymore
to eigenstates of the Hamiltonian, that is to stationary states, but they are called resonances.
As explained in Sec.~\ref{subsec:adsctac}, this corresponds to the case where the discrete state
[the first component of the ansatz (\ref{eq:ansatz_dimeron}), here with $\KK=\OO$] washes out in the continuum
[the last component of the ansatz (\ref{eq:ansatz_dimeron}), parameterized by the continuous variables
$\kk$, $\kk'$, $\qq$]. If one prepares the system in the complex energy state,
it will decay from that state into the continuum, with an exponential law
$\exp(-Gamma t)$ for the population, even though the system is isolated.
At times $\gg 1/\Gamma$, we expect that the impurity
momentum will be given by $\qq-\kk-\kk'$ as in the last component of (\ref{eq:ansatz_dimeron}) and
will have a broad probability distribution.

We therefore have to solve the integral equation~(\ref{eq:system}) 
with a complex energy-dependent kernel given in equation (\ref{eq:kernel}). 
Allowing $E$ to be complex in the function $F_{\kk'\kk\qq}(E,\OO)$ appearing in the kernel is straightforward, since apart from the pole at 
$E=0$ it is continuous over the whole complex plane. Instead $\alpha_{\kk\qq}(E,\OO)$ is a discontinuous function of $E$
and has a branch cut at most given by the interval $[0,+\infty)$ on the real axis as explained in the Appendix~\ref{app:integral}. Therefore for all $k>k_\mathrm{F}$ and $q<k_\mathrm{F}$ 
we perform an analytic continuation $\alpha^{\rm cont}_{\kk\qq}(E,\OO)$ through the branch cut from above
so that
\be
\label{eq:alphaC}
\lim_{\epsilon\to0^+} \left[\alpha_{\kk\qq}(E_0+i\epsilon,\OO)-
\alpha^{\rm cont}_{\kk\qq}(E_0-i\epsilon,\OO)\right]=0,
\ee
and in the Appendix~\ref{app:integral} we explicitly show how to satisfy this condition.

We then solve at $\PP=\OO$ the implicit equation (\ref{eq:implicit_dim}) for a complex $\Delta E_{\rm dim}$
with Newton's method \cite{newton_complexe}.
In Fig.~\ref{fig:complex-ener} we show a particular case for $r=1$ and $k_{\rm F}R_* = 1$, where the dimeronic energy becomes complex
as soon as it crosses the zero threshold, which corresponds to the lower border of the dimeron continuum. This situation
corresponds to the scenario of Fig.~\ref{fig:sketch}c.
Notice that the imaginary part showed in panel (b) is about two orders of magnitude smaller than its
corresponding real part in (a), which indicates a long lifetime of the dimeron. In addition, in the limit where $a\rightarrow 0^-$ we find that the real part
saturates to the Cooper's energy value $E_\mathrm{F}/r$ while the imaginary part vanishes:
\begin{eqnarray}
\mathrm{Re}\{\Delta E_\mathrm{dim}(\OO)\} &\stackrel{a \to 0^-}{\longrightarrow}& E_\mathrm{F}/r, \\
\mathrm{Im}\{\Delta E_\mathrm{dim}(\OO)\} &\stackrel{a \to 0^-}{\longrightarrow}& 0^-.
\end{eqnarray}
Even if not shown we find for the same parameters many other complex 
solutions to the implicit equation (\ref{eq:implicit_dim}) which have a larger real part; 
the larger the real part, the larger the absolute value of the imaginary part. 
Finally we have observed similar behaviors also for different
values of the Feshbach length, in particular also for $k_{\rm F}R_* = 0$, and different values of the mass ratio $r$.

%It is worth noting that in the limit where $a\rightarrow 0^-$ a real solution can be obtained using a simplified version of the dimeronic ansatz 
%\be
%\label{eq:simple_dimeron}
%\ket{\widetilde{\psi}_{\rm dim}(\OO)} = \sum'_{\kk',\kk,\qq} \eta_{\kk'\kk\qq} \hat{d}_{\qq-\kk-\kk'}^\dag \hat{u}_{\kk'}^\dag 
%\hat{u}_\kk^\dag \hat{u}_\qq  \ket{\mathrm{FS}:{N-1}},
%\ee
%which consists of setting $\eta=\eta_\kk=\eta_{\kk\qq}=0$ into Eq.~(\ref{eq:ansatz_dimeron}). Minimizing the expectation value of $H$ within
%the ansatz $(\ref{eq:simple_dimeron})$ gives the energy
%\be
%\widetilde{\Delta E}_{\rm dim}(\OO) = E_{\qq-\kk-\kk'}+\varepsilon_\kk+\varepsilon_{\kk'}-\varepsilon_\qq-E_{\rm F},
%\ee
%which attains its minimum value $\widetilde{\Delta E}_{\rm dim}(\OO) = 0$ when the vectors $\qq$, $\kk$ and $\kk'$ have all modulus equal to $k_{\rm F}$ 
%and are directed such that $\qq-\kk-\kk' = \OO$. In particular, this solution is slightly higher than the polaronic solution in the limit $a\rightarrow 0^-$, indicating that 
%in such a limit the ground state belongs to the polaronic branch.
% 

\subsection{Variational ansatz with $\ell=1$ at $\PP=\OO$}
\label{sec:strange}

\begin{figure}[htbp]
\begin{center}
\includegraphics[width=1.0\linewidth,clip=]{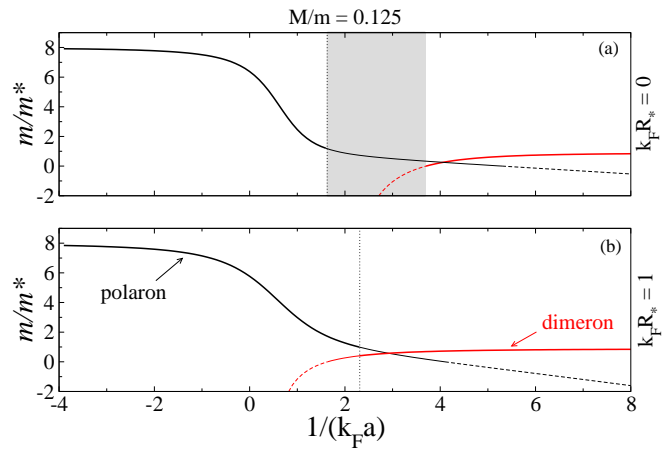}
\caption{(Color online) Polaron and dimeron inverse effective mass calculated for $r=0.125$ and for two different values of the Feshbach length: (a) $k_{\rm F}R_*=0$ and (b) $k_{\rm F}R_*=1$. The convention for the lines is 
as in Fig.~\ref{fig:meff}. The gray-shaded area indicates what we call the instability region, 
where the ground state of the system 
within our variational ansatz is dimeronic with a corresponding negative effective mass. }
\label{fig:meff0.125}
\end{center}
\end{figure}

In Fig.~\ref{fig:meff}, for a mass ratio $r=0.1505$,
it was observed that, in a narrow region of parameters (see the shaded
area), our $\PP=\OO$ and $\ell=0$ variational ground state is dimeronic with a {\sl negative} effective mass 
and is unstable. It can thus
not be the absolute ground state of the system.
By further decreasing the mass ratio, the instability region becomes larger as can be seen in Fig.~\ref{fig:meff0.125}
for $k_{\rm F} R_*=0$, whereas it is absent for $k_{\rm F}R_*=1$ as in Fig.~\ref{fig:meff}.

Within the instability region, there are three possible scenarios: (i) the absolute ground state of the system is 
at some non-zero momentum $\PP_0$ (as suggested by the negative effective mass in $\PP=\OO$), 
(ii) the absolute ground state has $\PP=\OO$ with a non-zero angular momentum,
(iii) the absolute ground state is of a nature not captured by the variational ansatz
(\ref{eq:ansatz_dimeron}).
Exploring the scenarios (i) and (iii) is beyond the scope of this paper.
We have unsuccessfully explored the scenario (ii) by solving the integral equation (\ref{eq:system}) 
when the function $\eta_{\kk\qq}$ has a total angular momentum $\ell=1$ (rather than $\ell=0$ elsewhere
in this paper), both for even and odd parities of that function,  as detailed in the Appendix~\ref{app:dim_l1}:
For $r=1/8$ and $r=0.1505$, with $k_{\rm F} R_*=0$, we found that, within the instability region,
the $\ell=1$ dimeronic solution has a {\sl larger} energy than the $\ell=0$ one.
The nature of the ground state of the system in this instability region remains an open question.

Finally we checked the polaron to dimeron crossing points with the $\ell=1$ ansatz for the dimeron and for the polaron, the latter being derived in the 
Appendix~\ref{app:polaron}. For all the crossings presented in Fig.~\ref{fig:crossings} we found 
that the $\ell=1$ ansatz provide a higher energy than the corresponding 
$\ell=0$ ansatz. 

\section{Analytical results in a non-trivial weakly attractive limit at $\PP=\OO$}
\label{sec:ariantwalapz}

The usual weakly attractive limit considered in the theory of quantum gases simply corresponds to $a\to 0^-$, the other parameters being fixed.
In our impurity problem on a narrow Feshbach resonance, this analytically tractable limit shall not give very interesting information,
as the leading order deviation from the ideal gas then depends on the scattering length $a$ only, the dependence on the Feshbach $R_*$
length being thus washed out. In particular, this can not explain the phase diagrams of Fig.~\ref{fig:crossing-3}.

A more interesting weakly attractive limit consists in taking $a\to 0^-$ while decreasing $\Lambda$
so that $R_* a$ is constant. In the expression of the two-body scattering amplitude,
one then sees that the $1/a$ and the $k_{\rm rel}^2 R_*$ terms are of the same order, for a fixed
relative momentum $k_{\rm rel}$ (of the order of $k_{\rm F}$), so that the scattering amplitude
$f_{k_{\rm rel}}$ tends to zero linearly in $a$ but still keeps a dependence on the relative momentum.
We shall now explore this non-trivial weakly attractive limit and show that it provides a quantitative
physical explanation to the numerical results.

\subsection{Results for the polaron}
\label{subsec:rftpazm}

Let us first sketch the usual weakly attractive limit $a\to 0^-$ for $R_*$ and $k_{\rm F}$ fixed, 
where we expect for our impurity problem that $\Delta E_{\rm pol}$ tends to zero (we are here at $\PP=\OO$).
In the integral appearing in the definition (\ref{eq:defD}) of $D_\qq(E)$ we then replace $E$ by zero, the integral remains 
finite for all values of $\qq$ in the Fermi sea, even for
$q\to k_{\rm F}$. Then, for $a\to 0^-$, one finds $D_\qq(E) \simeq 1/g$ which, upon
insertion in the implicit equation for $\Delta E_{\rm pol}$, reduces to the expected mean field result
\be
\label{eq:bmf}
\Delta E_{\rm pol} \sim \rho g,
\ee
where $\rho$ is the fermionic gas density. However all features due to the two-channel model
have been left out in this trivial limit.

In the non-trivial weakly attractive limit, we reduce the scattering length to $0^-$ and correspondingly decrease the interchannel
coupling $\Lambda$ to keep $a R_*$ fixed.
More precisely, we define for $a<0$
\be
s \equiv \frac{r}{1+r} (-a R_*)^{1/2} k_{\rm F},
\label{eq:defs}
\ee
and we take the limit $a\to 0^-$ for a fixed $s$. Assuming that $\Delta E_{\rm pol}$ tends to $0^-$ in that
limit, so that the integral appearing in the definition (\ref{eq:defD}) of
the function $D_\qq(\Delta E)$ remains bounded \cite{foot_pire_cas}, we have
\be
\label{eq:lim_gdq}
g D_{\qq}(\Delta E_{\rm pol}) \underset{a\to 0^-}{\to} 1- (sq/k_{\rm F})^2.
\ee
As $D_{\qq}$ has to be negative for all $q<k_{\rm F}$, this set of assumptions is consistent for $s<1$.
After evaluation of the integral over $\qq$ in the implicit equation 
(\ref{eq:Epol}) for $\Delta E_{\rm pol}$, we obtain for $s<1$:
\be
\Delta E_{\rm pol} \underset{a\to 0^-}{\sim} \frac{\hbar^2 k_{\rm F}^2}{m} \frac{1+r}{r} \frac{k_{\rm F} a}{\pi} \frac{1}{s^2} 
\left[\frac{1}{2s} \ln \frac{1+s}{1-s}-1\right].
\label{eq:epol_lead}
\ee
Note that, for $s\ll 1$, this reduces to the usual mean field result (\ref{eq:bmf}).
This calculation also allows to obtain the effective mass of the polaron in the non-trivial
weakly attractive limit: In Eq.~(\ref{eq:msgen}), one replaces $g D_{\qq}(\Delta E_{\rm pol})$ 
by its limit (\ref{eq:lim_gdq}) to obtain \cite{converge}
\be
\frac{M}{m^*_{\rm pol}}  \underset{a\to 0^-}{=}  1 + \frac{2 k_{\rm F} a}{3\pi}  
\frac{1+r}{r^2} \frac{s^2}{(1-s^2)^2} + O\left((k_{\rm F}a)^2\right).
\ee

What happens to the polaronic energy when $a\to 0^-$ for a fixed $s>1$? It is reasonable to assume
that $\Delta E_{\rm pol}$ has then a finite, non-zero and negative limit that depends on $s$. In this case
\be
g D_{\qq}(\Delta E_{\rm pol}) \underset{a\to 0^-}{\to} 1- s^2\left[(q/k_{\rm F})^2+\frac{1+r}{r} \frac{\Delta E_{\rm pol}}{E_{\rm F}}\right].
\label{eq:limDgen}
\ee
Let us now consider the implicit equation for $\Delta E_{\rm pol}$: 
Since $1/D_{\qq} \sim g$, one may expect that
$\Delta E_{\rm pol}$ tends to zero\ldots except if the integration over $\qq$ leads to a divergence
that compensates for the factor $g$, due to the vanishing of $D_{\qq}$ at the border of the Fermi surface. 
At this order of approximation, 
for $s>1$ the discrete state energy $\Delta E_{\rm pol}$ 
is the solution of $D_{k_{\rm F} \ve_z}(\Delta E,\OO)=0$.
It thus coincides with the lower border $\Delta E_{\rm pol}^{\rm cont}$
of the polaron continuum, for $\PP=\OO$,
see the discussion below Eq.~(\ref{eq:border_polaron}).
An expansion of $\Delta E_{\rm pol}^{\rm cont}$ in powers of
$k_{\rm F} a$ is readily obtained. To zeroth order
\be
\Delta E_{\rm pol}\underset{a\to 0^-}{\to} 
\Delta E_{\rm pol}^{\rm cont,(0)}
=E_{\rm F} \frac{r}{1+r} \left(\frac{1}{s^2}-1\right).
\label{eq:epol_lead_2}
\ee
To get the next order correction, in (\ref{eq:limDgen}) 
one now has to keep the additive term
$-\mu k_{\rm F}/(\pi^2\hbar^2)$ and the integral, but one can approximate
$\Delta E_{\rm pol}^{{\rm cont}}$ by its zeroth order approximation (\ref{eq:epol_lead_2}) in that
integral. This leads to a first order correction 
\be
\Delta E_{\rm pol}^{\rm cont,(1)} = -\frac{rE_{\rm F} }{1+r} \frac{k_{\rm F}a}{\pi s^2}
\left[2+\mathcal{F}\left(\frac{2}{1+r},\frac{1-r^2/s^2}{(1+r)^2},1\right)\right]
\ee
where the function $\mathcal{F}$ is defined in the Appendix \ref{app:integral}.

We can however be more precise and show that the discrete state is expelled 
from the continuum, as in the scenario of Fig.~\ref{fig:sketch}b,
by an exponentially small quantity. Expressing the energy difference
as
\be
\label{eq:epol_lead_3}
\Delta E_{\rm pol} - \Delta E_{\rm pol}^{\rm cont} = 
-E_{\rm F} \frac{r}{1+r} \delta,
\ee
with $\delta >0$,
it appears that the denominator $D_\qq(\Delta E_{\rm pol},\OO)$
does not exactly vanishes in $q=k_{\rm F}$, but it reaches a value $\approx \delta$
linearly in $k_{\rm F}-q$, so that the integral in the right-hand side of 
Eq.~(\ref{eq:Epol}) diverges as $g \ln(1/\delta)$.  Since the left-hand
side of Eq.~(\ref{eq:Epol}) is close to $\Delta E_{\rm pol}^{\rm cont}$,
which remains finite in the non-trivial weakly attractive limit,
we find that $\delta$ vanishes exponentially fast
for $g\to 0^-$. The final result is (we recall that $s>1$) \cite{lemme}:
\be
\delta \underset{a\to 0-}{\sim} A 
\exp\left[\pi\left(\frac{r}{1+r}\right)^2(s^2-1)/(k_{\rm F} a)\right],
\ee
where the prefactor $A$ can be determined analytically if necessary.
With the same reasoning applied to Eq.~(\ref{eq:msgen}), we conclude 
that, for $s>1$,  the inverse effective mass of the polaron 
exponentially rapidly tends to $-\infty$ in
the non-trivial weakly attractive limit:
\be
\frac{M}{m_{\rm pol}^*} \underset{a\to 0-}{\sim} -\frac{2}{3(1+r)\delta}.
\ee

We can also proceed with the analytical evaluation of the closed channel population in the polaronic branch.
For $a<0$, taking as independent parameters $(s, R_*)$ rather than $(a,R_*)$, the general
expression (\ref{eq:Ncc_def}) for the mean number of closed channel molecules reads
\be
N_{\rm cc} =  -\frac{m}{\hbar^2 k_{\rm F}^2} \frac{1+r}{r} s^3 \frac{\partial \Delta E}{\partial s}.
\label{eq:ncc_with_s}
\ee
From Eqs.(\ref{eq:epol_lead},\ref{eq:epol_lead_2}) we then find the simple results:
\bea
N_{\rm cc}^{\rm pol} \underset{a\to 0^-}{\to} 0 \ \ \ & \mbox{for} &\ \ \ s<1, \\
N_{\rm cc}^{\rm pol} \underset{a\to 0^-}{\to} 1 \ \ \ & \mbox{for} &\ \ \ s>1.
\eea

Finally, we physically explain why the results for the polaron depend
on the location of $s$ with respect to unity, using the picture of a discrete
state coupled to a continuum exposed in subsection \ref{subsec:adsctac}.
According to Eq.(\ref{eq:cfhanlbc}), expressed in the non-trivial weakly
attractive limit, we immediately 
find that the lower border of the polaron continuum
is zero for $s<1$ and is negative for $s>1$.
So the physical situation for $s<1$ corresponds to Fig.~\ref{fig:sketch}a,
and for $s>1$ it corresponds to Fig.~\ref{fig:sketch}b.

\subsection{Results for the dimeron in the Cooper approximation}
\label{subsec:rfdica}

The same non-trivial weakly attractive limit as in the previous subsection may be applied to the dimeronic case.
As shown by the numerics, the Cooper approximation
is accurate in the relevant regime, in practice close to the crossing point between
the polaronic and the dimeronic branches. We shall provide an analytical justification of this fact
in subsection \ref{subsec:walfda}. For $\PP=\OO$, the implicit equation for the dimeron (that we now call cooperon
since this is in the Cooper approximation) is 
\be
D_{\OO}(\Delta E_{\rm Cooper}+E_{\rm F},\OO)=0.
\label{eq:impli_cooper}
\ee
As long as $\Delta E_{\rm Cooper} < E_{\rm F}/r$, the integral appearing in the expression of $D_\OO$ 
is well defined. Multiplying Eq.(\ref{eq:impli_cooper}) by $g$ and taking the limit $g\to 0^-$ for a fixed
$s$, one gets
\be
1-s^2 \frac{1+r}{r} \left(1+\frac{\Delta E_{\rm Cooper}}{E_{\rm F}}\right) =0,
\ee
which leads to the prediction
\be
\Delta E_{\rm Cooper} \underset{a\to 0^-}{\to} \Delta E_{\rm Cooper}^{(0)}\equiv E_{\rm F} \left[\frac{r}{1+r} \frac{1}{s^2} - 1\right].
\label{eq:ecoop_leading}
\ee
This is meaningful for  $\Delta E_{\rm Cooper}^{(0)} < E_{\rm F}/r$ that is for $s>r/(r+1)$
[otherwise $\Delta E_{\rm Cooper}\to E_{\rm F}/r$ exponentially fast].
It is then found that, in this weakly attractive limit, the polaronic and the dimeronic energy curves
cross when $s=[r/(1+r)]^{1/2}$, at a vanishing energy in that limit, which leads to the prediction
\be
\label{eq:ntwalr}
\left(\frac{1}{k_{\rm F} a}\right)_c \underset{R_*\to +\infty}{\sim} - \frac{r}{1+r} k_{\rm F} R_*.
\ee
This reproduces quite well the slopes of the numerical results of Fig.~\ref{fig:crossing-3}. To get a prediction for the intercept,
one simply has to calculate $\Delta E_{\rm Cooper}$ to next order 
in $k_{\rm F} a$. In (\ref{eq:impli_cooper})
one now has to keep the additive term $-\mu k_{\rm F}
/(\pi^2\hbar^2)$ and the integral, but one can approximate
$\Delta E_{\rm Cooper}$ by its zeroth order approximation (\ref{eq:ecoop_leading}) in that integral.
This leads to a first order correction
\begin{multline}
\Delta E_{\rm Cooper}^{(1)} = -\frac{\hbar^2 k_{\rm F}^2}{m} 
\frac{r}{1+r} \frac{k_{\rm F} a}{\pi}  \frac{1}{s^2} \\
\times 
\left[1 -\frac{r}{2(1+r)s}\ln \frac{s\frac{1+r}{r}+1}{s\frac{1+r}{r}-1}\right].
\label{eq:ecoop_first_corr}
\end{multline}
By equating $\Delta E_{\rm Cooper}^{(0)}+\Delta E_{\rm Cooper}^{(1)}$ to the leading order expression 
Eq.(\ref{eq:epol_lead}) for the polaron energy we obtain for the crossing point in the weakly attractive limit \cite{dsc}:
\begin{widetext}
\begin{multline}
\!\!\!
\left(\frac{1}{k_{\rm F} a}\right)_c \underset{R_*\to +\infty}{=} -\frac{r}{1+r} k_{\rm F} R_* 
+\frac{2}{\pi} \left\{
1-\left(\frac{1+r}{r}\right)^2 \!\!\!
%\phantom{\ln\frac{\left(\frac{r}{r+1}\right)^{1/2}}{\left(\frac{r}{r+1}\right)^{1/2}}}
%\right. %\\
%\left.
+\frac{1}{2} \left[\left(\frac{1+r}{r}\right)^{5/2} \!\!\!
-\left(\frac{r}{1+r}\right)^{1/2}\right] 
\ln\frac{1+\left(\frac{r}{1+r}\right)^{1/2}}{1-\left(\frac{r}{1+r}\right)^{1/2}}\right\} %\\
+O\left(\frac{1}{k_{\rm F} R_*}\right).
\label{eq:kfac_analy}
\end{multline}
\end{widetext}
This asymptotic prediction is in very good agreement with the numerical results, already
for moderately large values of $k_{\rm F} R_*$, see Fig.~\ref{fig:crossing-3}. For a mass ratio 
$M/m=1$, the asymptotic prediction for $k_{\rm F} R_*=0$ exceeds the full numerical calculation 
by $\simeq 3\%$ only, which is of course very fortuitous, if one realizes that the corresponding value
of $(k_{\rm F} a)_c$ is positive while the asymptotic result was obtained in a limit where $a\to 0^-$\ldots
Note that the intercept, that is the constant term in Eq.(\ref{eq:kfac_analy}), is positive for
all mass ratios, it diverges as $2/(3\pi r)$ for $r\to 0$ and it vanishes as
$3[\ln(4r)-4/3]/(\pi r)$ for $r\to +\infty$.

The closed channel population can also be accessed for the dimeronic branch in the Cooper
approximation, from the general formula (\ref{eq:ncc_with_s}).
For $s>r/(1+r)$ one finds that it tends to unity in the weakly attractive limit:
\be
N_{\rm cc}^{\rm Cooper} \underset{a\to 0^-}{\to} 1.
\ee

\subsection{Weakly attractive limit for the full dimeronic ansatz}
\label{subsec:walfda}

We now argue, in the previously considered weakly attractive limit ($a\to 0^-$ with $a R_*$ fixed),
and for $\PP=\OO$, that the energy $\Delta E_{\rm dim}$ corresponding to the full dimeronic ansatz
only weakly differs from the Cooper result $\Delta E_{\rm Cooper}$, in the sense that,
for fixed $a R_*$,
\be
\label{eq:deltaE}
\delta E \equiv \Delta E_{\rm dim}-\Delta E_{\rm Cooper} \underset{a\to 0^-}{=} O(a^2).
\ee
As a consequence, the expansions (\ref{eq:ecoop_leading},\ref{eq:ecoop_first_corr}) still 
apply for $\Delta E_{\rm dim}$, and the asymptotic result for the crossing point
(\ref{eq:kfac_analy}) is unaffected at the considered order.
For simplicity, we restrict to the regime where $\Delta E_{\rm Cooper} <0$, so that $\Delta E_{\rm dim}$
really exists as an energy minimum (and not as a complex pole in an analytic continuation,
see subsection \ref{subsec:continuation}).
This regime contains in particular the crossing point of the dimeronic and polaronic branches, as we have
seen in subsection \ref{subsec:rfdica}.

The idea is to perform a perturbative expansion, calculating $\delta E$ to leading
order in $a$, for $a\to 0^-$ with $s$ of Eq.(\ref{eq:defs}) fixed. Multiplying the integral equation (\ref{eq:system})  
by $g \alpha_{\OO\OO}(E)/\alpha_{\kk\qq}(E)$ for $E$ equal to the dimeronic relative energy,
we get the linear system:
\begin{widetext}
\begin{multline}
g \alpha_{\OO\OO}(\Delta E_{\rm dim})\, \eta_{\kk\qq} = 
\frac{g^2}{g \alpha_{\kk\qq}(\Delta E_{\rm dim})} \int'\!\!\frac{d^3q'}{(2\pi)^3} \frac{d^3k'}{(2\pi)^3} 
\frac{\eta_{\kk'\qq'}}{F_{\kk\OO\OO}(\Delta E_{\rm dim})F_{\kk'\OO\OO}(\Delta E_{\rm dim})} \\
+
\frac{g^2\alpha_{\OO\OO}(\Delta E_{\rm dim})}{g\alpha_{\kk\qq}(\Delta E_{\rm dim})}
\left[\int'\!\!\frac{d^3q'}{(2\pi)^3} \frac{\eta_{\kk\qq'}}{F_{\kk\OO\OO}(\Delta E_{\rm dim})}
-\int'\!\!\frac{d^3k'}{(2\pi)^3} \frac{\eta_{\kk'\qq}}{F_{\kk'\kk\qq}(\Delta E_{\rm dim})}\right].
\label{eq:ftwis}
\end{multline}
\end{widetext}
At the energy $\Delta E_{\rm Cooper}$, $g \alpha_{\OO\OO}$ vanishes by definition, so we linearize in $\delta E$, calculating
the derivative of $\alpha_{\OO\OO}$ with respect to the energy to leading order in the weakly attractive limit:
\be
g \alpha_{\OO\OO}(\Delta E_{\rm Cooper}+\delta E) \simeq -\frac{1+r}{r} \frac{s^2}{E_{\rm F}}\, \delta E.
\label{eq:lin_a00}
\ee
The left-hand side of (\ref{eq:ftwis}) is thus of order $(\delta E/E_{\rm F}) \eta_{\kk\qq}$.
In the right-hand side of (\ref{eq:ftwis}), it appears that $g \alpha_{\kk\qq}$ has a finite limit
for $g\to 0^-$ with fixed $s$ 
~\cite{pire}:
\be
 g \alpha_{\kk\qq}(\Delta E_{\rm Cooper}) \to \frac{1+r}{r} 
\frac{s^2}{E_{\rm F}} \left[\varepsilon_\kk-\varepsilon_\qq+\frac{\varepsilon_{\qq-\kk}}{1+r}\right].
\ee
Note that this limit is non-negative, and vanishes only for $\qq$ and $\kk$ having coinciding values on the
Fermi surface. Also the functions $F$ in the denominators of (\ref{eq:ftwis}) do not depend on $g$. Taking into account the linearization
(\ref{eq:lin_a00}), it appears that the first term in the right-hand side of (\ref{eq:ftwis}),
which is not proportional to $\delta E$, is of order $g^2$. On the contrary, the second term 
in that right-hand side is of order $g \delta E$, with $g\to 0^-$, and is negligible as compared to the left-hand side
of (\ref{eq:ftwis}), and thus as compared to the first term of the right-hand side.

\begin{figure}[tbp]
\begin{center}
\includegraphics[width=1.0\linewidth,clip=]{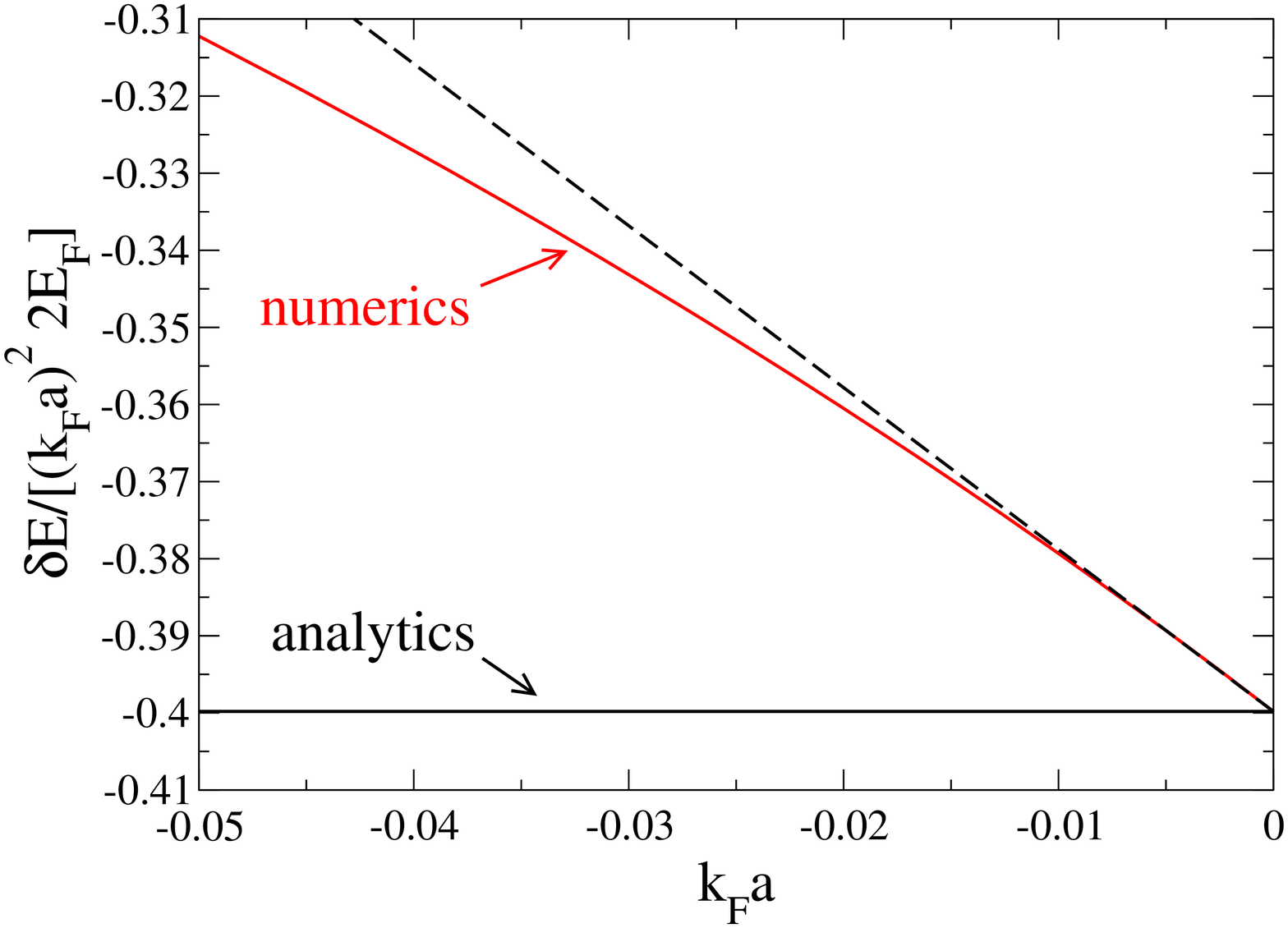}
\caption{(Color online) Dimeron-to-Cooperon energy difference (\ref{eq:deltaE}) divided by $(k_{\rm F}a)^2$ as a function of $k_{\rm F}a$, calculated for a mass ratio $r=6.6439$ at $s=s_c=[r/(1+r)]^{1/2}$
where $s$ is defined by Eq.~(\ref{eq:defs}).
To leading order in the non-trivial weakly attractive limit,
 $s=s_c$ corresponds to the polaron-to-dimeron crossing point. The horizontal line is the analytic Eq.~(\ref{eq:deltaEA}),
which is reached by the full numerical calculation only in the limit $a\to0^-$ (with $aR_*$ fixed). The dashed line is the slope of the linear terms in the numerical data.}
\label{fig:deltaE}
\end{center}
\end{figure}

To leading order in $g$, we finally have the eigenvalue problem
\begin{multline}
\delta E \, \eta_{\kk\qq} = - \left(\frac{g E_{\rm F}}
{\frac{1+r}{r} s^2 }\right)^2 
\left(\varepsilon_\kk-\varepsilon_\qq+\frac{\varepsilon_{\qq-\kk}}{1+r}\right)^{-1} \\
\times \int'\frac{d^3q'}{(2\pi)^3} \frac{d^3k'}{(2\pi)^3}\frac{\eta_{\kk'\qq'}}{F_{\kk\OO\OO}(\Delta E_{\rm Cooper}^{(0)})
F_{\kk'\OO\OO}(\Delta E_{\rm Cooper}^{(0)})},
\label{eq:eiv}
\end{multline}
where the zeroth-order approximation for the Cooper energy is given by (\ref{eq:ecoop_leading}).
This eigenvalue problem is solved by integrating over $\kk$ and $\qq$ after division by $F_{\kk\OO\OO}$, which leads to
the explicit expression for the leading deviation of the dimeron energy from the cooperon energy:
\begin{multline}
\label{eq:deltaEA}
\delta E \simeq - \left(\frac{g E_{\rm F}}
{\frac{1+r}{r} s^2 }\right)^2 \int'\!\!\!\frac{d^3q}{(2\pi)^3} \frac{d^3k}{(2\pi)^3}
\left(\varepsilon_\kk-\varepsilon_\qq+\frac{\varepsilon_{\qq-\kk}}{1+r}\right)^{-1} \\
\times \left(\frac{r}{1+r} \frac{E_{\rm F}}{s^2}-\frac{\hbar^2 k^2}{2\mu}\right)^{-2}.
\end{multline}
As promised, this vanishes as $g^2$ in the non-trivial weakly attractive limit.
Obviously, one also gets the corresponding leading order approximation for $\eta_{\kk\qq}$.
Note that our expression for $\delta E$ is negative, which was expected from the usual
argument that the variational space for the dimeronic ansatz contains the cooperonic ansatz \cite{convergence_issues}.
In Fig.~\ref{fig:deltaE} we successfully compare the energy difference (\ref{eq:deltaEA}) with the corresponding 
energy difference calculated numerically with the variational ansatz (\ref{eq:ansatz_dimeron}) and (\ref{eq:ansatz_Cooper}). 

To complete this analytical discussion of the dimeronic ansatz, we use the
discussion of subsection \ref{subsec:adsctac} to have a prediction on the
lower border of the dimeron continuum at $\PP=\OO$ in the non-trivial
weakly attractive limit. Evaluating the lower
border $\Delta E_{\rm pol}^{\rm cont} (\hbar\KK)$ 
of the polaron continuum for an arbitrary 
total momentum $\hbar \KK$ can be done 
by solving Eq.~(\ref{eq:border_polaron}) in that limit and minimizing
the root over $\qq$ inside the Fermi sea: One obtains either 
$\Delta E_{\rm pol}^{\rm cont} (\hbar\KK)=0$ for
small $s$, or
\be
\label{eq:bicplntif}
\Delta E_{\rm pol}^{\rm cont} (\KK) \to 
\frac{r}{1+r}(s^{-2}-1) +\frac{1}{1+r} \left[(K/k_{\rm F})^2 
-2 K/k_{\rm F}\right],
\ee
when $s$ is large enough for the right-hand side of (\ref{eq:bicplntif}) to
be negative. From Eq.~(\ref{eq:upper_dimeron}) we conclude that
\be
\Delta E_{\rm dim}^{\rm cont}(\OO) \le \min\left(0,
\frac{r}{1+r} \frac{1}{s^2} -1 \right) E_{\rm F}.
\ee
The lower border of the dimeron continuum is thus negative in the non-trivial
weakly attractive limit for 
$s>[r/(1+r)]^{1/2}$, which is the same range of $s$ leading to
a negative cooperon energy, see Eq.~(\ref{eq:ecoop_leading}).
For $s$ slightly above $[r/(1+r)]^{1/2}$,
the lowering of the dimeron continuum border pushes down the dimeron
discrete state energy, according to the scenario of
Fig.~\ref{fig:sketch}b, leading to its crossing with the polaron energy
(that remains vanishingly small for $s<1$ when $a\to 0^-$).

\section{Conclusion}
\label{sec:conclusion}

We have studied in detail the problem of a single impurity of mass $M$ interacting with a spatially homogeneous Fermi sea of same-spin-state fermions 
of mass $m$ on a $s$-wave narrow Feshbach resonance. 
As we have discussed, this problem belongs to
the general class of a discrete state coupled to a continuum.
Due to the narrowness of the Feshbach resonance there is, 
in addition to the $s$-wave scattering length,  
a second parameter characterizing the interaction,
namely the Feshbach length $R_*$.
Using simple variational ansatz limited to at most one pair of particle-hole excitations of the Fermi sea with Fermi momentum $k_{\rm F}$, 
we have determined for the polaronic and dimeronic branches the phase diagram between absolute ground state, local minimum, thermodynamically unstable regions (with negative effective mass), and regions of complex energies (with negative imaginary part). In particular,
we have numerically calculated the polaron-to-dimeron
crossing point $1/(k_{\rm F}a)_c$,  
where  $a$ is the $s$-wave scattering length, as a function of $R_*$ and 
for different values of the mass ratio $r=M/m$.
One of our main results is that the polaron-to-dimeron crossing point is shifted in the negative $a$ region for large $k_{\rm F} R_*$.
A physical explanation of this fact is given below.
Putting forward a non-trivial weakly attractive limit where $a\to0^-$ and $R_*\to\infty$, 
with the product $aR_*$ being constant, we could obtain analytical results for the polaronic and dimeronic branches and finally get the asymptotic expansion (\ref{eq:kfac_analy}) for the crossing point. 
We have calculated experimentally accessible properties of the system,
such as the effective mass both for the polaron  $m_{\rm pol}^*$ and for the dimeron $m_{\rm dim}^*$, as well as the closed-channel molecule population $N_\mathrm{cc}$ for the two branches.

For $M/m=0.1505$ we have found a small region of parameters where the ground state of the system {\sl restricted} to zero total linear and angular momenta is dimeronic with a negative effective mass $m_{\rm dim}^*<0$, therefore indicating
an instability of the system.
Extension of the dimeronic variational calculation to the angular momentum $\ell=1$ (with still $\PP=\OO$)  has not allowed to clarify the nature of the absolute ground state of the system in this instability region.

We now propose a physical interpretation of the shift towards negative values of $a$ of the polaron-to-dimeron crossing point for a narrow Feshbach resonance.
First, in a broad Feshbach resonance, described with a single channel model, a two-body bound state in free space can exist only if $a>0$. 
A common wisdom %\cite{Leyronas} 
is that the presence of the Fermi sea makes it more difficult for the impurity to form a two-body bound state with one of the fermions.
Such a bound state at zero momentum is indeed a spatially localized object formed by the impurity and a fermion by pairing of their
momenta $\pm \hbar\kk$ over some interval of values of $k$.
The presence of the Fermi sea limits the momentum available for the considered fermion to be larger than $\hbar k_\mathrm{F}$.

In a narrow Feshbach resonance, described with a two-channel model, there is from the start a molecular state in the closed
channel and the issue is to understand intuitively why it does not immediately give rise to a two-body bound state between the impurity
and a fermion. In short, due to the interchannel coupling $\Lambda$, the molecular state can dissociate into a continuum of open channel states where the impurity and the fermion have opposite momenta, if the molecular energy is above the dissociation threshold
of the open channel.  A naive application of this argument in free space, would lead to the wrong conclusion that there is a two-body bound state iff $E_{\rm mol}<0$. But the closed channel molecular state experiences
a strongly negative ``Lamb shift" due the coupling $\Lambda$ with the vacuum fluctuation in the open channel \cite{heff}, 
so that $E_{\rm mol}$ is renormalized into:
\be
\label{eq:emol_tilde}
\tilde{E}_{\rm mol} = E_{\rm mol} - \Lambda^2\int\frac{d^3k}{(2\pi)^3} \chi^2(\kk) \frac{2\mu}{\hbar^2k^2}.
\ee 
According to Eq.~(\ref{eq:Emol}), $\tilde{E}_{\rm mol}$ has the simple expression $\tilde{E}_{\rm mol}=-\Lambda^2/g$. The condition that there is a two-body bound state in free space iff
$\tilde{E}_{\rm mol}<0$ gives the correct condition that $a>0$. In the presence of the Fermi sea in the open channel, there are two changes: 
(i) the effective dissociation threshold is now the Fermi energy $E_{\rm F}$ \cite{mure_refl}, 
and (ii) in the ``Lamb shift" 
(\ref{eq:emol_tilde}) one has to restrict the integration over $\kk$ to $k> k_{\rm F}$, which makes that shift slightly less negative.  Therefore there is a two-body bound state
iff the resulting $\tilde{E}_{\rm mol}$ is less than $E_{\rm F}$, that is
\be
\label{eq:resultat_simple}
\frac{1}{k_{\rm F}a} > \frac{2}{\pi}-\frac{M}{m+M}k_{\rm F}R_*.
\ee
In the right-hand side of that equation, the first term originates from
the modification of the ``Lamb shift" by the Fermi sea, and the second one
from the rise of the dissociation threshold by the Fermi energy.
Remarkably, this simple reasoning perfectly agrees with the non-trivial weakly attractive limit lowest order result~(\ref{eq:ntwalr}), where $k_{\rm F} R_*\to \infty$  \cite{note_careful}.
In the opposite limit of a broad Feshbach resonance, it reproduces the expected result that the two-body bound state cannot form unless $1/k_{\rm F} a$ is positive and large enough~\cite{not_always_true}.

\subsection*{Acknowledgments}
We warmly thank  X. Leyronas, C. Lobo, S. Giraud, A. Sinatra, F. Werner for fruitful discussions. The group of Y.C. is a member of IFRAF.  We acknowledge financial support from the ERC Project FERLODIM N.228177. C. Trefzger acknowledges financial support from IFRAF and CNRS.

\subsection*{Note}

While completing this work we became aware of related studies \cite{Massignan_etroite,Asiatique_etroite}.

\appendix
\section{Implicit equation for the polaron energy}
\label{app:polaron}

Here we derive the implicit equation (\ref{eq:Epol}) for the polaron energy $\Delta E_{\rm pol}(\PP)=E_{\rm pol}(\PP) - E_{\rm FS}(N)$. Equating
to zero the first order derivatives of $\bra{\psi_\mathrm{pol}(\PP)}\hat{H}-E_{\rm pol}(\PP) \ket{\psi_\mathrm{pol}(\PP)}$ 
with respect to the complex conjugates of the variational parameters leads to the following set of coupled equations
\be
\left[\Delta E_{\rm pol}(\PP) - E_\KK\right] \phi = \frac{\Lambda}{\sqrt{V}} \sum_{\qq\in \mathrm{FS}_N} 
\chi[\mu(\frac{\qq}{m}-\frac{\KK}{M})] \phi_\qq 
\ee
\be
[\Delta E_{\rm pol}(\PP) + \varepsilon_\qq - \varepsilon_\kk -E_{\KK+\qq-\kk}] \phi_{\kk\qq} = \frac{\Lambda}{\sqrt{V}} 
\chi[\kk-\frac{\mu}{M}(\KK+\qq)]\phi_\qq 
\label{eq:couplee_phiq}
\ee
\begin{multline}
\left[\Delta E_{\rm pol}(\PP)+\varepsilon_\qq  -\frac{\varepsilon_{\KK+\qq}}{1+r}- E_\mathrm{mol}\right] \phi_\qq =  
\frac{\Lambda}{\sqrt{V}}\\
\times \Big\{\chi[\mu(\frac{\qq}{m}-\frac{\KK}{M})] \phi  
+ \sum_{\kk\notin \mathrm{FS}_N} \chi[\kk-\frac{\mu}{M}(\KK+\qq)] \phi_{\kk\qq}\Big\}.
\label{eq:coupled_phikq}
\end{multline}
This can be reduced to a single equation for the variational parameter $\phi_\qq$: 
From the first equation we eliminate $\phi$ and from the second equation we 
eliminate $\phi_{\kk\qq}$, using the fact that the energy differences cannot vanish for $\Delta E_{\rm pol} <0$.
We then divide the third equation by $\Lambda^2$ and we replace the internal energy of the molecule
by its expression Eq.(\ref{eq:Emol}). We take the thermodynamic limit, replacing the discrete sums over $\qq$ and $\kk$
by integrals. We split the integral in Eq.(\ref{eq:Emol}) in two bits, an integral for $k<k_{\rm F}$ and an integral for $k>k_{\rm F}$.
The bit with $k>k_{\rm F}$ can be collected with the integral over $\kk$ resulting from Eq.(\ref{eq:coupled_phikq}): This allows to directly
take the infinite momentum cut-off limit, where $\chi\to 1$, without activating a divergence. Using Eq.(\ref{eq:Rstar}) we finally obtain
\be
\label{app:Epol0}
D_\qq[\Delta E_{\rm pol}(\PP),\PP] \phi_\qq = \frac{1}{\Delta E_{\rm pol}(\PP) - E_\KK}  \int' \!\!\! \frac{d^3q'}{(2\pi)^3} \phi_{\qq'},
\ee
where $D_\qq[E(\PP),\PP]$ is given in Eq.(\ref{eq:defD}). A non-trivial solution of Eq.(\ref{app:Epol0}) may be easily found by 
making the change of variable
\be
\label{app:varphi}
\varphi = \int' \!\!\! \frac{d^3q}{(2\pi)^3} \phi_{\qq}.
\ee
Then, if $D_\qq[\Delta E_{\rm pol}(\PP),\PP] \neq 0$ for all $q<k_{\rm F}$  we can divide Eq.(\ref{app:Epol0}) by
$D_\qq$ and integrate on $\qq$ over the Fermi sphere.
Dividing the resulting equation by $\varphi$ under the assumption $\varphi\neq0$, we recover the scalar implicit
equation for the polaron energy (\ref{eq:Epol}). 
For $\PP=\OO$ the corresponding solution for $\phi_\qq$ has zero total angular momentum
$\ell=0$ since $D_\qq$ is then rotationally invariant.
Note that the assumption $\varphi\ne 0$ necessarily fails if one rather looks for a solution $\phi_\qq$ with $\ell >0$.
Then Eq.(\ref{app:Epol0}) reduces to
\be
\label{app:Epol2}
D_\qq[\Delta E_{\rm pol}(\OO),\OO] = 0,
\ee
for some $\qq$ inside the Fermi sea.
As expected, and as checked numerically, the solution to Eq.(\ref{app:Epol2}) has always a higher energy 
than the solution with $\ell=0$ \cite{raison}.

\section{Explicit form of the integrals in  $D_{\qq}$, in $\alpha_{\kk\qq}$  and in $\alpha'_{\kk\qq}$}
\label{app:integral}

First we calculate the integral 
\be
\label{eq:defI}
I_{\qq}(E,\PP)= \int' \!\!\frac{d^3k'}{(2\pi)^3}
\left[\frac{1}{E_{\KK+\qq-\kk'}+\varepsilon_{\kk'}-\varepsilon_\qq-E}-\frac{2\mu}{\hbar^2k^{'2}} \right]
\ee
that appears in the definition (\ref{eq:defD}) of the function
$D_\qq(E,\PP)$. This is equivalent to calculating the function $\alpha_{\kk\qq}(E,\PP)$ according to
(\ref{eq:defalpha}), and the reciprocal property holds since
$D_\qq(E,\PP)= \alpha_{\OO\qq}(E-E_{\rm F},\PP)$.
To be sure that the integral is well defined, one has to check that the energy denominator
in Eq.(\ref{eq:defI}) remains positive for all values of $\KK$ (over the whole Fourier space),
of $\qq$ (such that $q<k_{\rm F}$) and of $\kk$ (such that $k>k_{\rm F}$). We thus impose the following
constraint on the energy $E$:
\be
E < \inf'_{\KK,\qq,\kk}  E_{\KK+\qq-\kk'}+\varepsilon_{\kk'}-\varepsilon_\qq = 0,
\ee
where the prime on the infimum symbol recalls the constraints $q<k_{\rm F}$ and $k>k_{\rm F}$.
Note that the condition $E<0$ is not restrictive since $E$ stands in practice for energy differences that
are negative in the ground state of the system.

The expansion of $E_{\KK+\qq-\kk'}$ in powers of $k'$ reveals that it is convenient to
use spherical coordinates $(k',\theta',\phi')$ of polar axis given by the direction
of $\KK+\qq$. The integration over the azimuthal angle $\phi'$ is straightforward, and
after the change of variables $x'=\cos \theta'$ and $\kappa'=k'/k_{\rm F}$ we are left with
\begin{multline}
I_{\qq}(E,\PP)= \frac{2\mu k_{\rm F}}{(2\pi\hbar)^2}\int_1^{+\infty} \!\!\! d\kappa' \kappa'^2 \\ 
\int_{-1}^1 \!\!\! dx' \Big[\frac{1}{\kappa'^2-A_{\qq}(\PP)x'\kappa' + B_{\qq}(E,\PP)}-\frac{1}{\kappa'^2}\Big],
\label{eq:Iinterm}
\end{multline}
where we have introduced
\bea
A_{\qq}(\PP)     &=& \frac{2\mu}{M}\frac{|\KK+\qq|}{k_{\rm F}} \\
B_{\qq}(E,\PP) &=& \frac{2\mu}{\hbar^2k_{\rm F}^2}[E_{\KK+\qq}-\varepsilon_\qq-E].
\eea
After integration over $x'$ this leads to integrals of the type
$\int d\kappa' \kappa' \ln(\kappa'^2 \pm a\kappa' + b),$
which have to be solved separately 
for the case $4b>a^2$ and $4b<a^2$. The final result is given by
\be
I_{\qq}(E,\PP)=\frac{-2\mu k_{\rm F}}{(2\pi\hbar)^2} 
\mathcal{F}\left[A_{\qq}(\PP),B_{\qq}(E,\PP),1\right],
\label{eq:I_vs_calf}
\ee
where we have introduced the function that vanishes for $c\to +\infty$:
\begin{multline}
\mathcal{F}(a,b,c) = -c+\frac{2c^2+2b-a^2}{4a}\ln\frac{c^2+ac+b}{c^2-ac+b} \\
\!\!\!\!- \left\{ \begin{array}{ll}
\!\!\! \frac{\sqrt{4b-a^2}}{2} \left(\arctan\frac{2c+a}{\sqrt{4b-a^2}} + \arctan\frac{2c-a}{\sqrt{4b-a^2}}-\pi \right) & \!\!\! \textrm{if $4b>a^2$} \\\\
\!\!\! \frac{\sqrt{a^2-4b}}{4} \left( \ln\frac{2c+a+\sqrt{a^2-4b}}{2c+a-\sqrt{a^2-4b}} + \ln\frac{2c-a+\sqrt{a^2-4b}}{2c-a-\sqrt{a^2-4b}} \right) & \!\!\!  \textrm{if $4b<a^2$}
\end{array} \right.
\label{eq:calF_expli}
\end{multline}

In the second part of this appendix, we calculate the function $\alpha'_{\kk\qq}(E,\PP)$ defined in (\ref{eq:defalphaprime}).
Since it is simply the derivative of $\alpha_{\kk\qq}(E,\PP)$ with respect to $P_z$, one is reduced, by virtue
of relation (\ref{eq:defalpha}), to the derivative of $D_{\qq}(E,\PP)$  with respect to $P_z$: 
\be
\partial_{P_z} D_{\qq}(E,\PP) = -\frac{\mu^2 R_*}{\pi \hbar^3} \frac{K_z+q_z}{m+M} + \partial_{P_z} I_{\qq}(E,\PP).
\ee
From the chain rule for derivatives applied to (\ref{eq:I_vs_calf}):
\begin{multline}
\partial_{P_z} I_{\qq}(E,\PP) = -\frac{\mu^2(K_z+q_z)}{\pi^2\hbar^3 M k_{\rm F}}
\\  \times \Big(\frac{k_{\rm F}}{|\KK+\qq|} \partial_a \mathcal{F} + \partial_b \mathcal{F}
\Big)[A_\qq(\PP),B_\qq(E,\PP),1].
\end{multline}
Note that $\partial_b \mathcal{F}$ directly gives access to $\partial_E D_\qq(E,\PP)$, 
and thus to $\partial_E\alpha_{\kk\qq} (E,\PP)$, which is useful
for a Newton search of the polaronic and the dimeronic energies. An explicit calculation of the partial
derivative of (\ref{eq:calF_expli}) with respect to $a$ gives
\begin{multline}
\partial_a \mathcal{F}(a,b,c) = \frac{c}{a}-\Big(\frac{c^2+b}{2a^2}+\frac{1}{4}\Big)\ln\frac{c^2+ac+b}{c^2-ac+b} + \\
\left\{ \begin{array}{ll}
\!\!\! \frac{+a}{2\sqrt{4b-a^2}} \!\! \left(\arctan\frac{2c+a}{\sqrt{4b-a^2}} + \arctan\frac{2c-a}{\sqrt{4b-a^2}}-\pi \right) & \!\!\!  \textrm{if $4b>a^2$} \\\\
\!\!\! \frac{-a}{4\sqrt{a^2-4b}} \!\! \left( \ln\frac{2c+a+\sqrt{a^2-4b}}{2c+a-\sqrt{a^2-4b}} + \ln\frac{2c-a+\sqrt{a^2-4b}}{2c-a-\sqrt{a^2-4b}} \right) & \!\!\!  \textrm{if $4b<a^2$}.
 \end{array} \right.
\end{multline}
The derivative of (\ref{eq:calF_expli}) with respect to $b$ gives
\begin{multline}
\partial_b \mathcal{F}(a,b,c) = \frac{1}{2a} \ln\frac{c^2+ac+b}{c^2-ac+b}+ \\
\left\{ \begin{array}{ll}
\!\!\! \frac{-1}{\sqrt{4b-a^2}} \!\! \left(\arctan\frac{2c+a}{\sqrt{4b-a^2}} + \arctan\frac{2c-a}{\sqrt{4b-a^2}}-\pi \right) & \!\!\!  \textrm{if $4b>a^2$} \\\\
\!\!\! \frac{+1}{2\sqrt{a^2-4b}} \!\! \left( \ln\frac{2c+a+\sqrt{a^2-4b}}{2c+a-\sqrt{a^2-4b}} + \ln\frac{2c-a+\sqrt{a^2-4b}}{2c-a-\sqrt{a^2-4b}} \right) & \!\!\!  \textrm{if $4b<a^2$}.
 \end{array} \right.
\end{multline}

In the third part of this appendix, we allow the energy to be complex $E\in \mathbb{C}$ and analytically continue the integral $I_\qq(E,\PP)$ defined in (\ref{eq:I_vs_calf}).
Taking the standard branch cut for the logarithm and square root functions of a complex variable, i.e. the interval $(-\infty,0]$ on the real axis, then the branch cut of $I_\qq(E,\PP)$ 
is given by the interval $[0,+\infty)$ on the real axis. We continue
$I_\qq(E,\PP)$ from above over this branch cut:
Given $a,c \in \mathbb{R}$ this is equivalent to analytically continue
$\mathcal{F}(a,b,c)$ as a function of $b\in \mathbb{C}$ so that
\be
\label{eq:condition-F}
\lim_{\epsilon\to0^+} \left[\mathcal{F}(a,b_0-i\epsilon,c) - \mathcal{F}(a,b_0+i\epsilon,c) \right] = 0,
\ee
for all $b_0=\mathrm{Re}(b)\in \mathbb{R}$. To satisfy the condition (\ref{eq:condition-F}) one has to calculate the same limits on each function appearing in $\mathcal{F}$
but in practice there are only two types of functions: The first one is $\ln(c^2 - ac + b)$ which can be discontinuous 
when $\mathrm{Re}(c^2 - ac + b)<0$:
\be
\label{eq:limit-log}
\lim_{\epsilon\to0^\pm} \ln(c^2 - ac + b_0 + i\epsilon) = \ln|c^2 - ac + b_0| \pm i\pi.
\ee
We then substitute $\ln(c^2 - ac + b)$ with the continuous function
\begin{multline}
\label{eq:smalf}
f(a,b,c) = \\ 
\left\{ \begin{array}{ll}
\ln(c^2 - ac + b) & \textrm{if} \;\; \mathrm{Re}(c^2 - ac + b)>0\\\\
\ln(-c^2 + ac - b) - i\pi & \textrm{if} \;\; \mathrm{Re}(c^2 - ac + b)<0, 
\end{array} \right.
\end{multline}
which satisfies $f(a,b_0-i\epsilon,c)=f(a,b_0+i\epsilon,c)$ when $\epsilon~\to~0^+$. Notice that we have conventionally chosen Eq.~(\ref{eq:smalf}) such
as to concord with the minus sign limit of Eq.~(\ref{eq:limit-log}) in the relevant region.
The second one is $\ln\left(2c-a \pm \sqrt{a^2-4b}\right)$ which in turns has two cases. When $\mathrm{Re}(a^2-4b)>0$, similarly
to the previous case we substitute $\ln\left(2c-a \pm \sqrt{a^2-4b}\right)$ with the continuous function
\begin{multline}
h_\pm(a,b,c) = \\ 
\left\{ \begin{array}{ll}
\!\!\! \ln\left(2c \! - \! a \! \pm \! \sqrt{a^2\!-\!4b}\right) & \! \textrm{if} \; \mathrm{Re}\left(2c \! - \! a \! \pm \! \sqrt{a^2\!-\!4b}\right)\!>\!0\\\\
\!\!\! \ln\left(a\!-\!2c\! \mp \! \sqrt{a^2\!-\!4b}\right)\! \pm i\pi & \! \textrm{if} \; \mathrm{Re}\left(2c\! -\! a\! \pm \!\sqrt{a^2\!-\!4b}\right)\!<\!0, 
\end{array} \right.
\end{multline}
which for $\epsilon~\to~0^+$ satisfies $h(a,b_0-i\epsilon,c)=h(a,b_0+i\epsilon,c)$. Instead when $\mathrm{Re}(a^2-4b)<0$ we
simply substitute $\sqrt{a^2-4b}$ with $i\sqrt{4b-a^2}$ and no other arrangement has to be done in the logarithmic functions.

\section{Calculation of the polaron and dimeron effective masses}
\label{app:dim-meff}

To calculate analytically the polaron effective mass, we simply expand the implicit equation
(\ref{eq:Epol}) up to second order in $\PP=\hbar \KK$, which requires in particular an expansion of
the function $D_\qq[\Delta E_{\rm pol}(\PP),\PP]$ up to second order in $\PP$. After integration over
the direction of $\qq$ in (\ref{eq:Epol}), terms that are linear in $\KK$ vanish for symmetry reasons and, after
division by $K^2$,  one is left with
\begin{multline}
\!\!\!\!\!\!\!\!\!\left(\frac{\hbar^2}{2m^*_{\rm pol}}\!-\!\frac{\hbar^2}{2M}\right) \!
{\Bigg\{1\!+\!\!\int'\!\!\! \frac{d^3q}{(2\pi)^3} 
\frac{1}{\mathcal{D}_\qq^2}
\left[\frac{\mu^2 R_*}{\pi\hbar^4}\!+\!\!\! \int'\!\!\!\frac{d^3k'}{(2\pi)^3} \frac{1}{F_{\kk'\qq}^2}
\right]\!\!\Bigg\}}\\
\!=\!\!\int'\!\!\!\frac{d^3q}{(2\pi)^3} \frac{1}{\mathcal{D}_\qq^2} 
\Bigg[
\frac{F_\qq^2}{3 \mathcal{D}_\qq}
-\frac{\mu^3 R_*}{2\pi\hbar^2M^2}
-\!\int'\!\!\! \frac{d^3k'}{(2\pi)^3} \frac{\hbar^4 (\kk'-\qq)^2}{3 M^2 \, F_{\kk'\qq}^3} \Bigg]
\label{eq:msgen}
\end{multline}
where we have introduced the short-hand notation $\mathcal{D}_\qq=D_\qq[\Delta E_{\rm pol}(\OO),\OO]$,
and we have defined the scalar quantity $F_{\kk'\qq}\equiv E_{\qq-\kk'}+\varepsilon_{\kk'}-\varepsilon_\qq-\Delta E_{\rm pol}(\OO)$
and the vectorial quantity (parallel to $\qq$ for symmetry reasons):
\be
\label{eq:defFq}
\mathbf{F}_\qq \equiv -\frac{\mu^2 R_*\,\qq }{\pi(m+M)\hbar^2} + \int'\frac{d^3 k'}{(2\pi)^3} \frac{\hbar^2(\kk'-\qq)/M}{F_{\kk'\qq}^2}.
\ee

One possibility to calculate $m^*_{\rm dim}$ for the dimeron is to
solve Eq.(\ref{eq:implicit_dim}) for a few values of $\PP\in[\OO,\PP_\mathrm{max}]$ and then extrapolate $m^*_{\rm dim}$ from Eq. (\ref{eq:quadratic}).
However, this procedure is computationally very costly and is not safe from numerical errors, 
in particular due to the arbitrary choice of $\PP_\mathrm{max}$.
Instead, we use a perturbative approach in which we take the limit $\PP \rightarrow \OO$ in the operator $M[E(\PP),\PP]$. 
Thus, a second order expansion gives
\begin{multline}
\label{eq:expansionM}
M[E(\PP),\PP] \simeq M[E(\OO),\OO] + \frac{P^2}{2m^*_{\rm dim}} \partial_E M[E(\OO),\OO] \\
+ \sum_i P_i \partial_{P_i} M[E(\OO),\OO]  + \frac{1}{2} \sum_{i,j} P_i P_j \partial_{P_i} \partial_{P_j} M[E(\OO),\OO],
\end{multline}
where the indices $i,j$ run over the three orthogonal components $(P_x,P_y,P_z)$ of $\PP$, and we consider the operators in (\ref{eq:expansionM}) 
proportional to $\PP$ to be small perturbations of the operator $M[E(\OO),\OO]$. At $\PP=\OO$, the unperturbed eigenvalue equation is given by
\be
\label{eq:unperturbed}
M[E(\OO),\OO] \ket{v_n} = \lambda_n[E(\OO),\OO] \ket{v_n},
\ee
and the shift to $\lambda_n[E(\OO),\OO]$ induced by the perturbations is then 
\begin{multline}
\label{eq:shift}
\lambda_n[E(\PP),\PP] \simeq \lambda_n[E(\OO),\OO] + \sum_{s=1}^3 \delta\lambda_n^{(s)}[E(\OO),\PP],
\end{multline}
where each contribution is calculated using first and/or second order perturbation theory as follows
\be
\delta\lambda_n^{(1)}[E(\OO),\PP] = \frac{P^2}{2m^*_{\rm dim}} \bra{v_n} \partial_E M[E(\OO),\OO] \ket{v_n},
\ee
\begin{multline}
\delta\lambda_n^{(2)}[E(\OO),\PP] = \bra{v_n} \sum_i P_i \partial_{P_i} M[E(\OO),\OO] \ket{v_n} \\
+ \sum_{m \neq n} \frac{|\bra{v_n} \sum_i P_i \partial_{P_i} M[E(\OO),\OO] \ket{v_m}|^2}{\lambda_n[E(\OO),\OO] - \lambda_m[E(\OO),\OO]},
\end{multline}
\be
\delta\lambda_n^{(3)}[E(\OO),\PP] = \bra{v_n} \frac{1}{2} \sum_{i,j} P_i P_j \partial_{P_i} \partial_{P_j} M[E(\OO),\OO] \ket{v_n}.
\ee
At $\PP=\OO$, the dimeronic energy $E(\OO)$ is found by setting to zero the minimal eigenvalue
of the operator $M[E(\OO),\OO]$, namely
\be
\lambda_0[E(\OO),\OO]=0,
\ee
which implies $\lambda_0[E(\PP),\PP]=0$ for a finite $\PP$, thus the effective mass is readily obtained from the implicit equation
\be
\label{eq:implicitmeff}
\sum_{s=1}^3 \delta\lambda_0^{(s)}[E(\OO),\PP] = 0.
\ee
Specifically, for the eigenvector $\ket{v_0}$ we use an ansatz of zero total angular momentum
\be
\label{eq:ansatz}
\braket{\kk,\qq}{v_0} = f(k,q,\theta),
\ee
where $\theta$ is the angle between the vector $\kk$ and $\qq$.
Without loosing in generality we can assume the vector $\PP$ to be oriented along the $\ve_z$ axis, i.e. $\PP = P \, \ve_z$, and taking into account 
the symmetry of the ansatz (\ref{eq:ansatz}) the effect of the perturbations on $\lambda_0[E(\OO),\OO]$ is then reduced to calculate 
\begin{eqnarray}
\delta\lambda_0^{(1)}[E(\OO),\PP] = \frac{P^2}{2m^*_{\rm dim}} \bra{v_0} \partial_E M[E(\OO),\OO] \ket{v_0}
\end{eqnarray}
\begin{multline}
\delta\lambda_0^{(2)}[E(\OO),\PP] = -P^2\sum_{m \neq 0} \frac{|\bra{v_0} \partial_{P_z} M[E(\OO),\OO] \ket{v_m}|^2}{\lambda_m[E(\OO),\OO]}
\end{multline}
\begin{eqnarray}
\delta\lambda_0^{(3)}[E(\OO),\PP] = \frac{P^2}{2}\bra{v_0} \frac{\partial^2}{\partial P_z^2} M[E(\OO),\OO] \ket{v_0}.
\end{eqnarray}
The effective mass of the dimeron is then easily obtained from Eq. (\ref{eq:implicitmeff}), and we have
\begin{multline}
\label{eq:meff}
\frac{1}{m^*_{\rm dim}} = \Big\{2\sum_{m \neq 0} \frac{|\bra{v_0} \partial_{P_z} M[E(\OO),\OO] \ket{v_m}|^2}{\lambda_m[E(\OO),\OO]} \\
-  \bra{v_0} \frac{\partial^2}{\partial P_z^2} M[E(\OO),\OO] \ket{v_0} \Big\}/\bra{v_0} \partial_E M[E(\OO),\OO] \ket{v_0}.
\end{multline}
While the operator $M^{(1)}[E(\OO),\OO] = \partial_E M[E(\OO),\OO]$ is of straightforward calculation, the terms in bracket are more complicated and in what follows we 
explain the detailed calculation.

{\bf First term in (\ref{eq:meff})}{ --- }
The kernel of the operator $M^{(2)}[E(\OO),\OO]=\partial_{P_z} M[E(\OO),\OO]$ is given by the expression
\begin{multline}
\mathcal{M}^{(2)}[E,\kk,\qq,\kk',\qq'] = -(2\pi)^6\delta(\kk-\kk')\delta(\qq-\qq')\alpha_{\kk\qq}'(E,\OO) \\
-\frac{\hbar}{M} \Big\{\frac{(2\pi)^3\delta(\kk-\kk')}{F_{\kk\OO\OO}^2(E,\OO)}k_z  + \frac{(2\pi)^3\delta(\qq-\qq')}{F_{\kk'\kk\qq}^2(E,\OO)}(q_z-k_z-k_z')  \\
+ \Big[\Big(\frac{k_z}{F_{\kk\OO\OO}}+\frac{k_z'}{F_{\kk'\OO\OO}}\Big)(\alpha_{\OO\OO}F_{\kk\OO\OO}F_{\kk'\OO\OO})^{-1}\Big](E,\OO) \Big\},
\end{multline}
where 
\be
\label{eq:defalphaprime}
\alpha_{\kk\qq}'(E,\PP) = \partial_{P_z} \alpha_{\kk\qq}(E,\PP)
\ee
can be evaluated with the formulas of appendix \ref{app:integral}.

Using the completeness relation $1=\sum_n \ket{v_n}\bra{v_n}$ the first term in the right hand side of Eq. (\ref{eq:meff}) may be written as
\begin{multline}
\label{eq:scalar}
\sum_{m \neq 0} \frac{|\bra{v_0} M^{(2)}[E(\OO),\OO] \ket{v_m}|^2}{\lambda_m[E(\OO),\OO]} \\
=  \bra{v_0} M^{(2)}[E(\OO),\OO]  \frac{1}{M[E(\OO),\OO] } M^{(2)}[E(\OO),\OO] \ket{v_0}.
\end{multline}
It is not difficult to check that $\ket{x[E(\OO)]}=M^{(2)}[E(\OO),\OO] \ket{v_0}$ may be written as
\be
\label{eq:source}
\braket{\kk,\qq}{x[E(\OO)]} = \frac{k_z}{k} u_x[k,q,\theta,E(\OO)] + \frac{q_z}{q} v_x[k,q,\theta,E(\OO)].
\ee
Given that $\theta$ is the angle between $\kk$ and $\qq$, Eq. (\ref{eq:source}) is evidently an odd function of $(\kk,\qq)$, 
and as explained in Ref.~\cite{Castin01} has total angular momentum 
$\ell=1$. Using the ansatz (\ref{eq:ansatz}), the functions appearing in Eq. (\ref{eq:source}) are then given by:
\begin{multline}
\label{eq:ux}
u_x(k,q,\theta,E) = \frac{\hbar}{M}\Big\{-\frac{k}{4\pi^2} \int_0^{k_{\rm F}} \!\!\!\!\!\! dq'q'^2 \!\!\! \int_{-1}^1\!\!\!\!\!\! dx' \frac{f(k,q',x')}{F_{\kk\OO\OO}^2(E,\OO)} \\
- \frac{k}{8\pi^4}\int_{k_{\rm F}}^\infty \!\!\!\!\!\! dk'k'^2 \!\!\! \int_0^{k_{\rm F}} \!\!\!\!\!\! dq'q'^2 \!\!\! \int_{-1}^1\!\!\!\!\!\! dx' \frac{f(k',q',x')}{(\alpha_{\OO\OO}F_{\kk\OO\OO}^2F_{\kk'\OO\OO})(E,\OO)} \\
+ \frac{1}{4\pi^2} \int_{k_{\rm F}}^\infty \!\!\!\!\!\! dk'k'^2 \!\!\! \int_{-1}^1\!\!\!\!\!\!  dx' \frac{f(k',q,x')}{[b_0^2(E)-b_1^2]^{3/2}} \\ 
\times \Big[\frac{\sqrt{1-x'^2}}{\sqrt{1-x^2}}b_1k' -b_0(E)k\Big] \Big\} -\alpha_{\kk\qq}'(E,\OO) ,
\end{multline}
and respectively
\begin{multline}
\label{eq:vx}
v_x(k,q,\theta,E) =  \frac{\hbar}{M}\frac{1}{4\pi^2} \int_{k_{\rm F}}^\infty \!\!\!\!\!\! dk'k'^2 \int_{-1}^1\!\!\!\!\!\!  dx' \frac{f(k',q,x')}{[b_0^2(E)-b_1^2]^{3/2}} \\
\times \Big[b_0(E) (q-k'x') -\frac{\sqrt{1-x'^2}}{\sqrt{1-x^2}}b_1k'x\Big] -\alpha_{\kk\qq}'(E,\OO),
\end{multline}
where as usual $x=\cos\theta$. Where appropriate we have substituted 
\be
\label{eq:Fdecomposition}
F_{\kk'\kk\qq}(E,\OO) =  b_0(E) + b_1\cos\phi,
\ee
where 
\begin{eqnarray}
b_0(E) &=& E +E_{\rm F} - \frac{\hbar^2}{2\mu}(k^2+k'^2) + \frac{\hbar^2q^2}{2}\Big(\frac{1}{m}-\frac{1}{M}\Big) \nonumber \\
&+& \frac{\hbar^2}{M}(\kk + \kk')\qq - \frac{\hbar^2}{M}kk'xx',
\end{eqnarray}
and respectively
\be
b_1 = -\frac{\hbar^2}{M}kk'\sqrt{1-x^2}\sqrt{1-x'^2},
\ee
and we have performed the integration in $\phi$ \cite{help01}.

Since the operator $M[E(\OO),\OO]$ has a vanishing eigenvalue $\lambda_0[E(\OO),\OO]=0$
it is not possible to invert it, but we can overcome the problem by solving
\be
\label{eq:system00}
M[E(\OO),\OO] \ket{y} = \ket{x[E(\OO)]}.
\ee
Notice that since the source term $\ket{x[E(\OO)]}$ has odd parity and total angular momentum $\ell=1$, it is easy to check with (\ref{eq:kernel}) that also $\ket{y}$ must have odd parity 
and unit angular momentum. Following Ref.~\cite{Castin01}, this reduces the unknown $\ket{y}$ to be of the form
\be
\braket{\kk,\qq}{y} = \frac{k_z}{k} u_y(k,q,\theta) + \frac{q_z}{q} v_y(k,q,\theta),
\ee
where $\theta$ is the angle between the vectors $\kk$ and $\qq$. Writing this ansatz into Eq. (\ref{eq:system00}) leads to an implicit equation for the energy $E(\OO)$, in the unknown functions $u_y(k,q,\theta)$ and $v_y(k,q,\theta)$:
\begin{multline}
\label{app:system}
\frac{k_z}{k} \left\{U_k + V_k -u_x\right\}[k,q,\theta,E(\OO)] \\
+ \frac{q_z}{q} \left\{U_q + V_q-v_x\right\}[k,q,\theta,E(\OO)] = 0,
\end{multline}
where
\begin{multline}
\label{app:Ukdef}
U_k(k,q,\theta,E) = \frac{1}{4\pi^2}\int_0^{k_{\rm F}} \!\!\!\!\!\! dq'q'^2\int_{-1}^1\!\!\!\!\!\! dx'\frac{u_y(k,q',x')}{F_{\kk\OO\OO}(E,\OO)} \\
+\frac{1}{4\pi^2}\int_{k_{\rm F}}^\infty \!\!\!\!\!\! dk' k'^2 \int_{-1}^1\!\!\!\!\!\! dx' \frac{z_+}{\sqrt{b_0^2(E)-b_1^2}}\frac{\sqrt{1-x'^2}}{\sqrt{1-x^2}}u_y(k',q,x') \\
-\alpha_{\kk\qq}(E,\OO)u_y(k,q,\theta),
\end{multline}
\begin{multline}
\label{app:Vkdef}
V_k(k,q,\theta,E) = \frac{1}{4\pi^2}\int_0^{k_{\rm F}} \!\!\!\!\!\! dq'q'^2\int_{-1}^1\!\!\!\!\!\! dx' x'\frac{v_y(k,q',x')}{F_{\kk\OO\OO}(E,\OO)},
\end{multline}
\begin{multline}
\label{app:Uqdef}
U_q(k,q,\theta,E) = \frac{1}{4\pi^2}\int_{k_{\rm F}}^\infty \!\!\!\!\!\! dk' k'^2 \int_{-1}^1\!\!\!\!\!\! dx' x'\frac{u_y(k',q,x')}{\sqrt{b_0^2(E)-b_1^2}}\\
- \frac{x}{4\pi^2}\int_{k_{\rm F}}^\infty \!\!\!\!\!\! dk' k'^2 \int_{-1}^1\!\!\!\!\!\! dx' \frac{z_+}{\sqrt{b_0^2(E)-b_1^2}}\frac{\sqrt{1-x'^2}}{\sqrt{1-x^2}}u_y(k',q,x'),
\end{multline}
\begin{multline}
\label{app:Vqdef}
V_q(k,q,\theta,E) = \frac{1}{4\pi^2}\int_{k_{\rm F}}^\infty \!\!\!\!\!\! dk'k'^2\int_{-1}^1\!\!\!\!\!\! dx'\frac{v_y(k',q,x')}{\sqrt{b_0^2(E)-b_1^2}}, \\
-\alpha_{\kk\qq}(E,\OO)v_y(k,q,\theta).
\end{multline}
Once again $x=\cos\theta$, where appropriate we have made the substitution (\ref{eq:Fdecomposition}) and performed the azimuthal integration in $\phi$ \cite{help01}.

Therefore, for a given energy $E=E(\OO)$ we find the unknown functions $u_y(k,q,\theta)$ and $v_y(k,q,\theta)$ by setting to zero 
the two lines of Eq. (\ref{app:system}) separately since they are linearly independent \cite{help02}. Then the scalar product (\ref{eq:scalar}) is straightforward
and is given by the integral
\begin{multline}
\braket{x[E(\OO)]}{y} = \frac{1}{24\pi^4} \int_{k_{\rm F}}^\infty \!\!\!\!\!\! dkk^2 \!\!\! \int_0^{k_{\rm F}} \!\!\!\!\!\! dqq^2 \!\!\! \int_{-1}^1\!\!\!\!\!\! dx \\
\Big\{ u_x^*[k,q,\theta,E(\OO)]u_y(k,q,\theta) + v_x^*[k,q,\theta,E(\OO)]v_y(k,q,\theta) \\
+x\Big[u_x^*[k,q,\theta,E(\OO)]v_y(k,q,\theta) + v_x^*[k,q,\theta,E(\OO)]u_y(k,q,\theta)\Big]\Big\}.
\end{multline}

{\bf Second term in (\ref{eq:meff})}{ --- }
The second term in the right hand side of Eq. (\ref{eq:meff}) can be calculated by noting that,
due to rotational invariance,
$\langle v_0|\partial^2_{P_z}M[E(\OO),\OO]|v_0\rangle=
\frac{1}{3}\sum_{i = x,y,z} \langle v_0|\partial^2_{P_i}M[E(\OO),\OO]|v_0\rangle.$
The kernel of the operator $M^{(3)}[E(\OO),\OO]=\sum_i \frac{\partial^2}{\partial P_i^2} M[E(\OO),\OO]/3$ is then given by
\begin{multline}
\mathcal{M}^{(3)}[E,\kk,\qq,\kk',\qq'] = - (2\pi)^6 \delta(\kk-\kk^\prime) \delta(\qq-\qq^\prime) \alpha_{\kk\qq}''(E,\OO) \\
+\frac{(2\pi)^3\delta(\kk-\kk')}{F_{\kk\OO\OO}^2(E,\OO)}\Big[1+\frac{2\hbar^2}{3M}\frac{k^2}{F_{\kk\OO\OO}(E,\OO)} \Big] \frac{1}{M} \\
-\frac{(2\pi)^3\delta(\qq-\qq')}{F_{\kk'\kk\qq}^2(E,\OO)} \Big[1+ \frac{2\hbar^2}{3M} \frac{(\qq-\kk-\kk')^2}{F_{\kk'\kk\qq}(E,\OO)}\Big] \frac{1}{M} \\
-\Big\{ \Big[\frac{\alpha_{\OO\OO}''}{\alpha_{\OO\OO}} - \frac{1}{M}\Big(\frac{1}{F_{\kk\OO\OO}} + \frac{1}{F_{\kk'\OO\OO}}\Big) + \frac{2\hbar^2}{3M^2}\Big(\frac{k^2}{F_{\kk\OO\OO}^2} + \frac{k'^2}{F_{\kk'\OO\OO}^2}\Big) \\
+\frac{2\hbar^2}{M^2} \frac{\kk\cdot\kk'}{F_{\kk\OO\OO}F_{\kk'\OO\OO}} \Big]
\times(\alpha_{\OO\OO}F_{\kk\OO\OO}F_{\kk'\OO\OO})^{-1}\Big\}(E,\OO)
\end{multline}
where
\begin{multline}
\label{eq:defalphasecond}
\alpha_{\kk\qq}''(E,\PP) = \sum_i \frac{\partial^2}{3\partial P_i^2} \alpha_{\kk\qq}(E,\PP) = \frac{-\mu^2R_*}{\pi\hbar^4(M+m)} \\
-\frac{1}{M} \int' \!\!\frac{d^3k'}{(2\pi)^3}
\Big[\frac{1}{F_{\kk^\prime\kk\qq}^2(E,\PP)}+\frac{2\hbar^2}{3M} \frac{(\KK+\qq-\kk-\kk')^2}{F_{\kk^\prime\kk\qq}^3(E,\PP)}\Big].
\end{multline}
As previously mentioned, the prime on the integral sign means that the integral is restricted
to $q<k_{\rm F}$ and/or to $k>k_{\rm F}$. Eq. (\ref{eq:defalphasecond}) may be calculated from Eq. (\ref{eq:defalpha}) by using the simple relation
\begin{multline}
\alpha_{\kk\qq}''(E,\PP) = \frac{\mu^2R_*}{\pi\hbar^4}\Big(\frac{1}{M}-\frac{1}{m+M}\Big)\\
-\frac{1}{M}\partial_E \alpha_{\kk\qq}(E,\PP) - \frac{2}{3M^2}\partial_{1/M} \partial_E [\alpha_{\kk\qq}(E,\PP)]_{R_*=0}.
\end{multline}

\section{Dimeronic ansatz with $\ell=1$}
\label{app:dim_l1}
Here we solve the integral equation (\ref{eq:system}) for $\eta_{\kk\qq}$ having total angular momentum $\ell=1$, both for even and odd parity.  
Following Ref.~\cite{Castin01}, for an even function of total angular momentum $\ell=1$ we write the following ansatz
\be
\label{app:even_ansatz}
\eta_{\kk\qq} = \ve_z \cdot \frac{\kk \wedge \qq}{|\kk \wedge \qq|} f(k,q,\theta),
\ee
where $\ve_z$ is a unit vector along the $z$ direction, $\wedge$ indicates the vectorial product, $\theta$ is the 
angle between the vectors $\kk$ and $\qq$.
We insert the ansatz (\ref{app:even_ansatz}) into the integral equation (\ref{eq:system}), and apart from the diagonal term the only non-zero integral is given by
\be
\int' \!\!\! \frac{d^3k'}{(2\pi)^3} \frac{\eta_{\kk'\qq}}{F_{\kk'\kk\qq}(E,\OO)},
\ee
where $E=\Delta E_\mathrm{dim}(\OO)$ is the dimeronic energy at $\PP=\OO$. In spherical coordinates $(k',\theta',\phi')$ this integral can be easily solved  by performing the substitution (\ref{eq:Fdecomposition}), and by integrating over the azimuthal angle~\cite{help01} we arrive at the following integral equation
\begin{multline}
\label{app:int_even}
 \Big[\frac{1}{4\pi^2} \int_{k_\mathrm{F}}^\infty \!\!\! dk' k'^2 \int_{-1}^1 \!\!\! dx' \frac{z_+}{\sqrt{b_0^2(E)-b_1^2}} f(k',q,x') \\
- \alpha_{\kk\qq}f(k,q,x)\Big] \ve_z \cdot \frac{\kk \wedge \qq}{|\kk \wedge \qq|} = 0,
\end{multline}
where as usual $x'=\cos\theta'$.

For an odd function of total angular momentum $\ell=1$ we write the ansatz
\be
\label{app:odd_ansatz}
\eta_{\kk\qq} = \frac{k_z}{k} u_y(k,q,\theta) + \frac{q_z}{q} v_y(k,q,\theta),
\ee
where $\theta$ is the angle between $\kk$ and $\qq$. We have already encountered in the Appendix \ref{app:dim-meff} the procedure of inserting this ansatz into the integral equation (\ref{eq:system}), which leads to an integral equation for the dimeronic energy $E=\Delta E_\mathrm{dim}(\OO)$ at $\PP=\OO$ given by
\begin{multline}
\label{app:system_odd}
\frac{k_z}{k} (U_k + V_k)[k,q,\theta,E] \\
+ \frac{q_z}{q} (U_q + V_q)[k,q,\theta,E] = 0,
\end{multline}
where $U_k$, $V_k$, $U_q$ and $V_q$ are given in Eqs. (\ref{app:Ukdef},\ref{app:Vkdef},\ref{app:Uqdef},\ref{app:Vqdef}). We then get the integral equation in its
final form by setting to zero the two lines of Eq. (\ref{app:system_odd}) separately since they are linearly independent~\cite{help02}:
\be
\left\{ \begin{array}{l}
U_k[k,q,\theta,E] + V_k[k,q,\theta,E] = 0 \\
U_q[k,q,\theta,E] + V_q[k,q,\theta,E] = 0.
\end{array}  \right.
\ee


\begin{thebibliography}{99}
\bibitem{link}
One has $R_*=\hbar^2/(2\mu a_{\rm bg} \Delta B \partial_{B} E_{\rm mol})$,
where $a_{\rm bg}$ is the background scattering length
such that $a(B)=a_{\rm bg} [1-\Delta B/(B-B_0)]$. The derivative of $E_{\rm mol}$ with
respect to the magnetic field $B$ is taken at the location $B=B_0$ of the resonance.
\bibitem{dimer_free}
The energy of a dimer in free space may be obtained from a pole of the scattering amplitude. For any value of $R_*$, setting $k_\mathrm{rel} = iq_\mathrm{rel}$ 
one finds that  Eq. (\ref{eq:amplitude}) has a pole with $q_\mathrm{rel}$ real and positive if and only if $a>0$, with the binding energy of the dimer given by
\be
%\label{eq:def_dimer}
E_\mathrm{d} = -\frac{\hbar^2q_\mathrm{rel}^2}{2\mu} \ \ \mbox{with}\ \ q_\mathrm{rel}=\frac{2/a}{1+\sqrt{1+4R_*/a}}. \nonumber
\ee
\bibitem{convergence_uniforme}
To have a {\sl uniform} spectral convergence of the more complete two-channel model towards the narrow Feshbach
resonance model, one has to impose that $E_{\rm mol}$ tends to $+\infty$ in the limit
$a_{\rm bg}, b\to 0$ (otherwise the true ground state energy will tend to $-\infty$). 
E.g., for the two-channel model of \cite{Castin01}, where $\chi(\kk)=\exp(-k^2 b^2/2)$
is used as the cut-off function for both the closed-to-open channel coupling and the separable
open channel interaction, one has $\mu \Lambda^2/(2\pi\hbar^2 E_{\rm mol})= (a^{-1}-a_{\rm bg}^{-1})
/\{[a_{\rm bg}^{-1}-(b\sqrt{\pi})^{-1}] [a^{-1}-(b\sqrt{\pi})^{-1}]\}$. When $a_{\rm bg}\to 0$ and $b\to 0^+$,
for fixed $a$ and $\Lambda$, one finds that $E_{\rm mol}\to +\infty$ iff $b\sqrt{\pi}-a_{\rm bg}\to 0^+$.
\bibitem{nzt}
In this paper we restrict to zero temperature, so that we do not have to include
thermally generated particle-hole pairs. At finite temperature, one may expect that these thermal
pairs have a negligible effect when $T\ll T_{\rm F}=E_{\rm F}/k_{\rm B}$.
This expectation is confirmed (for $1/a=0$ and $R_*=0$) in \cite{Nascimbene}
where the effectively zero-temperature equation of state of a polaronic gas was measured.
\bibitem{help}
The equation (\ref{eq:implicit_dim}) actually has several solutions, even for $\PP=\OO$. 
To obtain the minimal energy solution, we found first that one should take for $\lambda$ the minimal
eigenvalue of $M[E,\OO]$.  
Second, we found it useful to calculate numerically $\Delta E_{\rm Cooper}$  
such that $\alpha_{\OO\OO}(\Delta E_{\rm Cooper},\OO)=0$.
As shown in \cite{Zwerger}, $\Delta E_{\rm Cooper}$ is the prediction for the dimeron coming from
a Cooper-type ansatz neglecting all particle-hole excitations in the Fermi sea
[that is setting $\eta_{\kk\qq}\equiv 0$ and $\eta_{\kk'\kk\qq}\equiv 0$ in
(\ref{eq:ansatz_dimeron})]. Since the variational ansatz explores a broader subspace,
one has $\Delta E_{\rm dim}(\OO) \leq \Delta E_{\rm Cooper}$, which is of crucial help in the numerical bracketing of
$\Delta E_{\rm dim}(\OO)$.
\bibitem{douvientelle}
To obtain this equation, we use Eqs.~(\ref{eq:couplee_phiq},\ref{eq:coupled_phikq}) with $\phi=0$. The hypothesis
of $\Delta E < 0$ ensures that the expression in between square brackets
in the left-hand-side of Eq.~(\ref{eq:couplee_phiq}) cannot vanish, which allows us to use 
Eq.~(\ref{eq:couplee_phiq}) to
eliminate $\phi_{\kk\qq}$ in terms of $\phi_{\qq}$ in Eq.~(\ref{eq:coupled_phikq}).
\bibitem{help_maximal}
For $\PP\ne\OO$ we found numerically that $D_\qq(\Delta E,\PP)$, for a fixed $\Delta E$, is maximal for $\qq$ on the Fermi surface with a direction opposite to $\PP$.
\bibitem{help_bound}
Contrarily to the polaronic case, a full minimization of the dimeronic ansatz with $\eta=0$ is made difficult
due to the coupling between $\eta_{\kk\qq}$ and $\eta_{\kk'\qq}$ arising the fermionic antisymmetry of $\eta_{\kk\kk'\qq}$.
\bibitem{limmef}
As $R_*\to \infty$, the quantity in between parenthesis that comes in factor of
$R_*$ in Eq.~(\ref{eq:defD}) has to vanish.
\bibitem{newton_complexe}
To evaluate the derivative of the eigenvalue $\lambda$ with respect to the energy
we use the generalized Hellmann-Feynman formula, which requires the calculation
of the left and right eigenvectors associated to $\lambda$. 
Since $M[E,\OO]$ is complex symmetric for complex $E$, the left eigenvector is simply
the complex conjugate of the right eigenvector.
\bibitem{foot_pire_cas}
For $q<k_{\rm F}$, the integral in Eq.~(\ref{eq:defD}) is finite for $\KK=0$ when
$\Delta E=0^-$ because the denominator of the integrand can not vanish.
This conclusion also holds in the worst case $q=k_{\rm F}$: The denominator vanishes
for $\kk'=\qq$ but the integral remains finite, as can be seen from integration
in spherical coordinates of polar axis along $\qq$.
\bibitem{converge}
One can directly approximate $\Delta E_{\rm pol}$ by $0^-$ in $F_{\kk'\qq}$ since
the integral over $\kk'$ and $\qq$ of $1/F^2_{\kk'\qq}$ still converges in Eq.~(\ref{eq:msgen}),
and the integral over $\kk'$ of $(\kk'-\qq)/F^2_{\kk'\qq}$ still converges in Eq.~(\ref{eq:defFq}).
\bibitem{lemme}
We used the lemma that, for a smooth non-negative
function $u(x)$ over the interval
$[0,1]$ that vanishes only in $x=0$ (where it vanishes linearly), one has
for $\delta\to 0^+$:
\begin{multline}
\int_0^1 \frac{dx}{u(x)+\delta}= 
\lim_{\eta\to 0^+} \Bigg[\int_\eta^{1} \frac{dx}{u(x)}-
\frac{\ln(1/\eta)}{u'(0)}\Bigg] \\ +\frac{\ln[u'(0)/\delta]}{u'(0)}
+O(\delta \ln \delta)
\nonumber
\end{multline}
\bibitem{pire}
In the worst case where $q=k_{\rm F}$, $k=k_{\rm F}$, $|\qq-\kk|=k_{\rm F}$, and setting $\Delta E_{\rm Cooper}=0^-$, 
it is found that the integral over $\kk'$ in the expression of $\alpha_{\kk\qq}(0^-)$
remains finite, even if the integrand diverges for $\kk'=\qq-\kk$. Its product with
$g$ thus vanishes for $g\to 0^-$.
\bibitem{convergence_issues}
To justify our perturbative expansion of (\ref{eq:ftwis}), we have 
addressed the convergence issues. First we have checked that 
the solution $\eta_{\kk\qq}$ of (\ref{eq:eiv}) is square integrable, 
even if it is not bounded. Second we have checked that one obtains 
a finite result when one contracts $\eta_{\kk \qq}$ \cite{point_technique}
with the neglected term in the right-hand side of Eq.~(\ref{eq:ftwis}).
Note that $F_{\kk'\kk\qq}$ in that term 
cannot vanish because of the supposed condition 
$\Delta E_{\rm Cooper}^{(0)} < = 0$.
Even the worst case $\Delta E_{\rm Cooper}^{(0)}=0$
is found to be convergent.
\bibitem{heff}
The ``Lamb shift" is calculated as usual with the $PGP$ formalism, where $G$ 
is the resolvent of the Hamiltonian and $P$ projects on the closed channel 
subspace \cite{CCT}. One then obtains an energy-dependent effective Hamiltonian 
that one evaluates at the energy $\Delta E=0$ to obtain the threshold 
condition for the  bound state.
\bibitem{mure_refl}
As a closed channel molecule of wavevector $\KK=\OO$ dissociates into a fermion
of wavevector $\kk$ and an impurity of wavevector $-\kk$, the final kinetic energy
is $\hbar^2 k^2/2\mu$ and one may conclude that the dissociation threshold in
presence of the Fermi sea is at an energy $\hbar^2 k_{\rm F}^2/2\mu$.
This reasoning ignores the subsequent scattering of the impurity on a particle
of the Fermi sea of wavevector $\qq$, that couples to a continuum with two fermions of wavevectors
$\kk$, $\kk'$ (with $k,k'> k_{\rm F}$), a hole of wave vector $\qq$ in the Fermi sea,
and the impurity of wavevector $-\kk-\kk'+\qq$. This continuum has an energy
threshold equal to $E_{\rm F}$.
\bibitem{note_careful}
A careful reader may argue that Eq.(\ref{eq:resultat_simple}) gives a condition not to cross the polaron energy but to have
a two-body bound state. However in the non-trivial weakly attractive limit the polaronic energy
vanishes at the crossing.
\bibitem{not_always_true}
As can be seen on the right-lower panel of Fig.~\ref{fig:phase-diagram} this expectation is not true for large mass ratio. The dashed line, which corresponds to $\Delta E_{\rm dim}(\OO)=0$, has
a negative intercept when $k_{\rm F}R_*$ tends to $0$.
\bibitem{details}
In practice, one uses the Lagrange multiplier method to take into account the fact
that the ansatz should have a constant norm, and one expresses the fact that the first
order derivatives with respect to the complex conjugates of the variational parameters
are zero. Here the Lagrange multiplier is simply the desired energy $E_{\rm pol}$ or
$E_{\rm dim}$ since the energy functional is quadratic.
\bibitem{raison}
A useful property is that $D_{\qq}[E]$ is an increasing function of $E$ [simply take the derivative
of Eq.(\ref{eq:defD}) with respect to $E$] that tends to $-\infty$ for $E\to -\infty$.
Solving Eq.(\ref{eq:Epol}) for $\PP=\OO$ amounts to solving $f(E)=0$, where
$f(E)=E-\int'\frac{d^3q}{(2\pi)^3} \frac{1}{\mathcal{D}_\qq(E)}$ is also an increasing function of $E$ 
that tends to $-\infty$ at $-\infty$.
A first case is that $\sup_{q<k_{\rm F}} D_\qq(E=0^-) < 0$. Then $f(0^-)>0$ and $f(E)$ has a negative root.
A second case is that there exists $E_*<0$ such that $\sup_{q<k_{\rm F}} D_\qq(E_*)=0$. For the parameters considered in this work,
we then always found numerically that $D_\qq(E_*)$
vanishes on the Fermi surface, linearly in $q-k_{\rm F}$, so that $f(E)$ diverges logarithmically in $E_*$ and $f(E)$ has a negative root below $E_*$.
\bibitem{help01}
Given $b_0,b_1 \in \mathbb{R}$ ($|b_0|> |b_1|$) and $n\in \mathbb{N}$, one faces integrals 
$I(n) = \int_0^{2\pi} d\phi \frac{\cos(n\phi)}{b_0+b_1\cos\phi}=2\pi z_+^{|n|} \mathrm{sign}(b_0)/\sqrt{b_0^2-b_1^2}$,
with $z_+=-b_1/[\mathrm{sign}(b_0)\sqrt{b_0^2-b_1^2}+b_0]$. Note that for numerical implementation $I(n)$ is in the right form,
since $I(n) \rightarrow 0 $ when $b_1 \rightarrow 0$. Furthermore, while $b_1<0$ by construction we have assumed
$b_0 < 0$.
\bibitem{help02}
In fact, except for a region of zero measure where the vectors $\kk$ and $\qq$ are parallel, one can always rotate 
along the axis of the $\kk$ vector to find $\qq\cdot\ve_z = 0$. Viceversa, rotating along the axis of the $\qq$ vector
leads to $\kk\cdot\ve_z = 0$.
\bibitem{point_technique}
A technical point is that, due to the non-hermitian character of the operator
appearing in Eq.~(\ref{eq:ftwis}), one has to contract the second term
in the right-hand side of that equation with the adjoint vector
of $\eta_{\kk\qq}$, which we found to be $\propto 1/F_{\kk\OO\OO}$,
since the change of function $\tilde{\eta}_{\kk\qq}= (g \alpha_{\kk\qq})^{1/2}
\eta_{\kk\qq}$ makes the integral operator hermitian in Eq.~(\ref{eq:ftwis}).
\end{thebibliography}
\end{document}